\documentclass[aps,prb,amsmath,amssymb,reprint,longbibliography]{revtex4-2}
\usepackage{bm}
\usepackage{graphicx}
\usepackage{amssymb}
\usepackage{amsmath}
\usepackage{eufrak}
\usepackage{physics}
\usepackage{color}
\usepackage[normalem]{ulem}
\usepackage[unicode=true,colorlinks=true,citecolor=blue,urlcolor=blue]{hyperref}

\newcommand{\mysection}[1]{\textcolor{blue}{\textit{#1}. }}
\newcommand{\nix}[1]{}

\begin{document}

\title{Linear photogalvanic effect in surface states of topological insulators}

\author{N.~V.~Leppenen}
\email{leppenen@mail.ioffe.ru}
%\affiliation{Ioffe Institute, 194021 St. Petersburg, Russia}
%
\author{L.~E.~Golub}
\affiliation{Ioffe Institute, 194021 St. Petersburg, Russia}

\begin{abstract}
Theory of the Linear photogalvanic effect is developed for direct optical transitions between surface states of three-dimensional topological insulators. The photocurrent governed by the orientation of the polarization plane of light and caused by the warping of the energy dispersion of two-dimensional carriers is calculated. 
It is shown that both the shift contribution caused by coordinate shifts of the particle wavepackets during the optical transitions and the ballistic contribution caused by interference of the optical absorption and scattering by disorder have generally the same order of magnitude. The ballistic contribution is present owing to electron-hole asymmetry of topological surface states and has a frequency dependence in contrast to the shift photocurrent. In the nonlinear in the light intensity regime appearing due to saturation of the direct optical transitions the ballistic contribution dominates. The nontrivial dependence of the photocurrent on the polarization plane orientation in the nonlinear regime appears solely due to the ballistic contribution. Our findings allow separating the ballistic and shift contributions to the Linear photogalvanic current in experiments.
\end{abstract}
\maketitle{}

\mysection{Introduction} Linear photogalvanic effect (LPGE) consists in an appearance of a photocurrent who's magnitude and direction are governed by an orientation of the light' linear polarization plane relative to crystallographic axes~\cite{Glass1974,Sturman_Fridkin_book,Ivchenko_book,Ganichev_book}. The electric current density $\bm j$ is related to the amplitude $\bm E$ of the light electric field $\bm E(t)=\bm E \exp(-i\omega t)+c.c.$ by the relation
\begin{equation}
\label{phenom_general}
j_\alpha = \chi_{\alpha\beta\gamma}E_\beta E_\gamma^*,
\end{equation}
where $\alpha, \beta, \gamma$ are Cartesian coordinates, and $\chi_{\alpha\beta\gamma}=\chi_{\alpha\gamma\beta}$ is a 3rd rank tensor symmetric in two last indices. It follows from this definition that the LPGE current exists in non-centro-symmetric media only: the tensor $\chi$ has the same symmetry as a piezoelectric tensor.
In some systems, the point symmetry allows for the LPGE current flow perpendicular to the linear polarization plane of light. This allowed one to interpret this current as a nonlinear Hall effect~\cite{NLHE_perspectives}.

LPGE has been observed in various condensed-matter systems as bulk semiconductors, quantum-well structures and topological insulators~\cite{andrianov1981linear,LPGE_GaN,LPGE_PRL_2014, LPGE_TI_photoioniz}. Theoretically, it was established four decades ago in 
Ref.~\cite{Belinicher_Ivchenko_Sturman_1982} that the current density at interband optical transitions is a sum of two contributions
\begin{equation}
\bm j = \bm j_{\rm ball} + \bm j_{\rm shift}.
\end{equation}
Here $\bm j_{\rm ball}$ is the so-called ballistic contribution (also sometimes called `injection' current~\cite{Sipe_SOR_2000}) caused by an asymmetry of photoexcitation, and $\bm j_{\rm shift}$ is the shift photocurrent caused by elementary shifts of electron wavepackets in the process of light absorption. Generally, these two contributions have the same order of magnitude, and only their sum has a physical meaning~\cite{Sturman_UFN_2020}. However, there are many theoretical works devoted to calculation of solely the shift contribution~\cite{shiftLPGE1,shiftLPGE2,shiftLPGE3}. 
At direct optical transitions between surface states of topological insulators, only the shift contribution has been calculated~\cite{Kim_Shift_2017}.

A deeper insight to the physics of LPGE is achieved by studies in the nonlinear in the intensity regime. Here, the tensor $\chi$ in Eq.~\eqref{phenom_general} depends on the light intensity $I \propto \abs{\bm E}^2$. Experimentally, high intensities needed for this nonlinear regime are available nowadays~\cite{EdgeGraphenePRB2021,LPGE_TI_photoioniz}. Such studies allow for determination of kinetic and band parameters of topological insulators which do not affect the LPGE current in the linear in $I$ regime~\cite{Nonlin_our_PRB_2022}.

%Phenomenologically, the 
The bulk structure of topological insulators of BiTe type is centrosymmetric with the non-symmorphic $D_{6h}$ point symmetry group with a non-trivial translation along the [111] direction. By contrast, their (111) surfaces have no such translation which lowers the symmetry of the surface to $C_{3v}$ point group. The latter is non-centrosymmetric with allowance for the LPGE~\cite{LPGE_PRL_2014,LPGE_TI_photoioniz}. 
The photocurrent direction is determined by the
orientation of the light polarization plane with respect to the crystallographic axes $(x,y)$ along and perpendicular to one of
the mirror-reflection planes of the $C_{3v}$ point group.

In this work, we develop a theory of LPGE for direct optical transitions 
between the valence and conduction surface states
of topological insulators. 
We demonstrate that  the LPGE photocurrent polarization dependence is given by
\begin{equation}
\label{phenom_C3v}
j_+ = \chi_1 E_-^2 + \chi_2 E_+^4,
\end{equation}
where the $\pm$ components of the vectors are defined as $a_\pm=a_x \pm ia_y$.
We calculate both ballistic and shift contributions to the LPGE current at arbitrary light intensities and
%We demonstrate 
show that these two contributions can be experimentally separated at high intensity.

\mysection{Model} The conduction and valence band states at the surface of the topological insulator are described by the low-energy Hamiltonian~\cite{shen2012topological}
\begin{equation}
\label{eq:H}
	{\cal H} = \epsilon (p)+v_0 [\bm \sigma\times \bm p]_z+\lambda \sigma_z \frac{p_+^3-p_-^3}{2i},
\end{equation} 
where $\bm p$ is the electron momentum,
the term $\epsilon(p)$ breaks the symmetry between conduction and valence bands, $\bm \sigma = (\sigma_x, \sigma_y)$ is the vector of Pauli matrices, and $v_0$ and $\lambda$ are the band structure parameters. The last term reflects the trigonal  $C_{3v}$ symmetry of the system~\footnote{A difference in the form of the last term of Eq.~\eqref{eq:H} in comparison with Refs.~\cite{hex_warping_L_Fu,li2014hexagonal} is due to another definition of the reflection planes of C$_{3v}$ point symmetry group.}. 
Note that this term yields a hexagonal warping correction to the electron energy 
$\propto \lambda^2$~\cite{hex_warping_L_Fu,li2014hexagonal}
 because the linear in $\lambda$ terms are forbidden by the time-inversion symmetry.
With account for the warping term in the first order, the wavefunctions have the form
\begin{equation}
\label{psi_cv}
\psi_{c,v} =\frac{1}{\sqrt{2}}\mqty[\pm ie^{-i\theta_{\bm p}/2} \qty(1\pm\Lambda_{\bm p}) \\ e^{i\theta_{\bm p}/2}(1\mp\Lambda_{\bm p})],
\quad
\Lambda_{\bm p}= {\lambda p^2\sin{3\theta_{\bm p}} \over 2 v_0},
\end{equation}
where $\theta_{\bm p}$ is the angle between $\bm p$ and the $x$ axis.  The energy dispersions in the bands are $\varepsilon_{c(v)\bm p} = \epsilon(p)\pm v_0 p$.

The electron-photon interaction Hamiltonian has the form 
$V  = {i e \over \omega} \bm E \cdot \bm \nabla_{\bm p} \mathcal H$.
Calculating the matrix element of a direct optical transition from the valence to the conduction surface band 
$V_{cv} = \left<c|V|v\right>$, we obtain up to the first order in $\lambda$
\begin{align}
\label{Vcv}
V_{cv}(\bm p)  = &-\frac{e E }{\omega} \biggl\{v_0\sin(\theta_{\bm p}-\alpha)\nonumber
\\
&+\frac{i\lambda p^2}{2} \qty[5\sin(2\theta_{\bm p}+\alpha)-\sin(4\theta_{\bm p}-\alpha)] \biggr\}.
\end{align}
Here linearly polarized light is assumed with $\alpha$ being the angle between the light polarization plane and 
%the $x$ axis.
the mirror-reflection plane $(zx)$.

\mysection{Ballistic contribution}
The ballistic LPGE current reperesents the contribution diagonal in the band indices:
\begin{equation}\label{eq:ball_general}
	\bm j_{\text{ball}} = e \sum_{\bm p} \qty(\bm v_{c\bm p} f_{c\bm p}+ \bm v_{v\bm p}f_{v\bm p}) .
\end{equation}
Here $\bm v_{n\bm p}=\bm \nabla_{\bm p}\varepsilon_{n\bm p}$ are the electron velocities, and $f_{n \bm p}$ are occupations  ($n=c,v$). The latter are the sums $f_0(\varepsilon_{c,v})\pm\Delta f_{c,v}$ where $f_0$ is the equilibrium Fermi-Dirac distribution and $\Delta f_{n \bm p}$ is the light-induced correction. 
The latter is found from the kinetic equation which in a steady-state has the form
\begin{equation}
\label{eq:kin:ur}
\text{St}_c[\Delta f_{c\bm p}] = \sum_{\bm p'} 
G_{\bm p \bm p'}
%W_{c\bm p,v\bm p'}
(1-\Delta f_{v\bm p'}-\Delta f_{c\bm p}).
\end{equation}
Here $\text{St}_c$ is the collision integral in the conduction band, and 
the generation rate is given by the Fermi Golden rule 
$G_{\bm p \bm p'}%W_{c\bm p,v\bm p'}
=(2\pi/\hbar)\abs{M_{c\bm p, v\bm p'}}^2 \delta(\varepsilon_{c\bm p}-\varepsilon_{v\bm p'}-\hbar\omega)$.
Kinetic equation for $\Delta f_{v\bm p'}$ is obtained from the above equation by interchanging $\bm p \leftrightarrow \bm p'$.

\begin{figure}[htp]
	\centering
	\includegraphics[width=\linewidth]{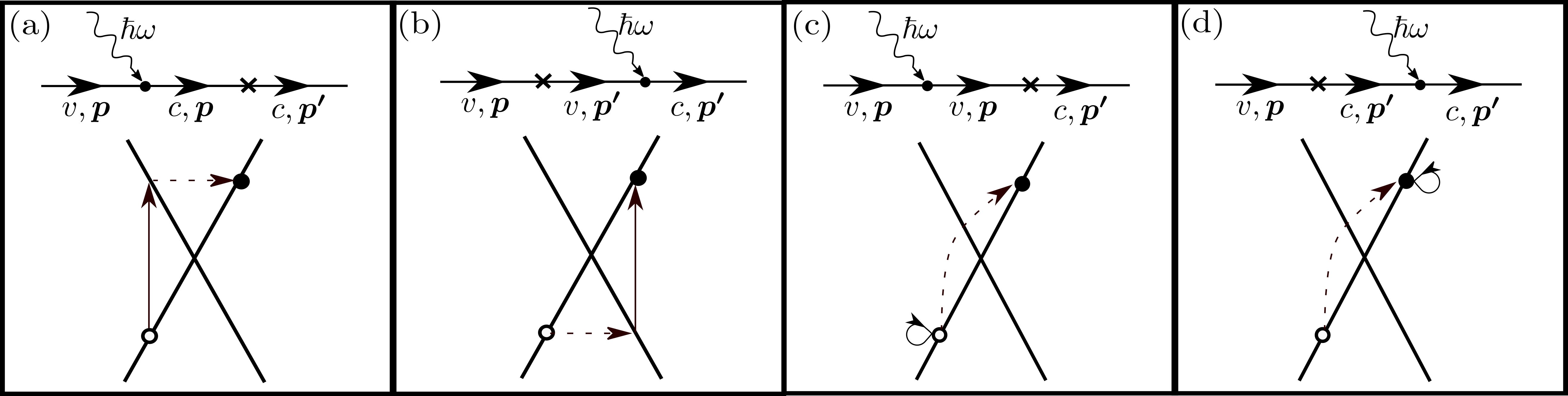} 
	\caption{Second order contributions to the matrix element of the transition $v\bm p \rightarrow c\bm p'$. The solid line shows interband (a,b) or intraband (c,d) optical transitions and dashed lines represent intraband (a,b) and interband (c,d) scattering.}
	\label{fig:2p2}
\end{figure}

The specific feature of the LPGE is that the lowest Born approximation, where $M_{c\bm p, v\bm p'}=\delta_{\bm p'\bm p}V_{cv}(\bm p)$, is insufficient for calculation of the photocurrent. Indeed, it follows from Eq.~\eqref{Vcv} that
$\abs{V_{cv}}^2$ has no $\lambda$-linear terms, and, hence, the corrections $\Delta f_{n \bm p}$ 
%calculated by the Fermi Golden rule 
can not 
%have $\lambda$-linear terms responsible for 
result in the photocurrent. However, it is possible to obtain the ballistic contribution to the LPGE current in the next to Born approximation with account for interference of the electron-photon interaction with disorder or phonon scattering processes.
The matrix element of disorder-assisted interband transition is a sum $M_{cv}=M_{cv}^{(2+2)}+M_{cv}^{(1+3)}$ of the terms of the 2nd order of the perturbation theory and those of the  1st and 3rd orders.
In the 2nd order we have for an indirect process:
\begin{equation}
M_{c\bm p', v\bm p}^{(2+2)}= \sum_{n} {U_{c\bm p',n}V_{nv}(\bm p) + V_{cn}(\bm p')U_{n,v\bm p} \over E_i-E_n + i0}.
\end{equation}
Here $U$ is the disorder potential, the initial total energy of the system $E_i=\varepsilon_{v\bm p} + \hbar\omega$, and summation is performed over intermediate states $n$ in the $c$- and $v$-bands.
The corresponding processes are depicted in Fig.~\ref{fig:2p2}.
The scattering matrix elements $U_{m\bm p',n\bm p}=U_{\bm p'\bm p}\left< m\bm p'|n\bm p\right>$ with $U_{\bm p'\bm p}$ being the Fourier-image of the disorder potential, and the intra-band electron-photon interaction matrix elements are given by 
$V_{nn}(\bm p)=ie\bm E \cdot \bm v_{n\bm p}/\omega$ 
%with $\bm v_{n\bm p}=\bm \nabla_{\bm p}\varepsilon_{n\bm p}$ 
($n=c,v$). 
The following relations could be obtained from the time-reversal symmetry 
 \begin{equation}\label{eq:PU_rel}
U_{v,\bm p';v,\bm p} = U_{c,\bm p;c,\bm p'},\qquad
	U_{c,\bm p';v,\bm p} = -U_{c,\bm p;v,\bm p'}.
\end{equation}

Analysis shows that the ballistic contribution from transitions shown in Fig.~\ref{fig:2p2} comes from the interference term 
\[ \abs{M_{c\bm p', v\bm p}^{(2+2)}} \propto 2\Re\qty[\qty(M^{(a)}+M^{(b)})^*\qty(M^{(c)}+M^{(d)})].
\] 
Furthermore,
only products of the real term from one process and the imaginary term from  another contribute to the current. 
Therefore the imaginary terms with $\delta$-functions should be taken from one of the denominators in the matrix elements $M^{(a,b,c,d)}$ . As a result, 
the energy conservation $\varepsilon_{c,\bm p'} =\varepsilon_{c,\bm p}$ or   $\varepsilon_{v,\bm p'} =\varepsilon_{v,\bm p}$ in the intermediate state appears -- see horizontal scattering arrows in Fig.~\ref{fig:2p2}.

\begin{figure}[htp]
	\centering
	\includegraphics[width=\linewidth]{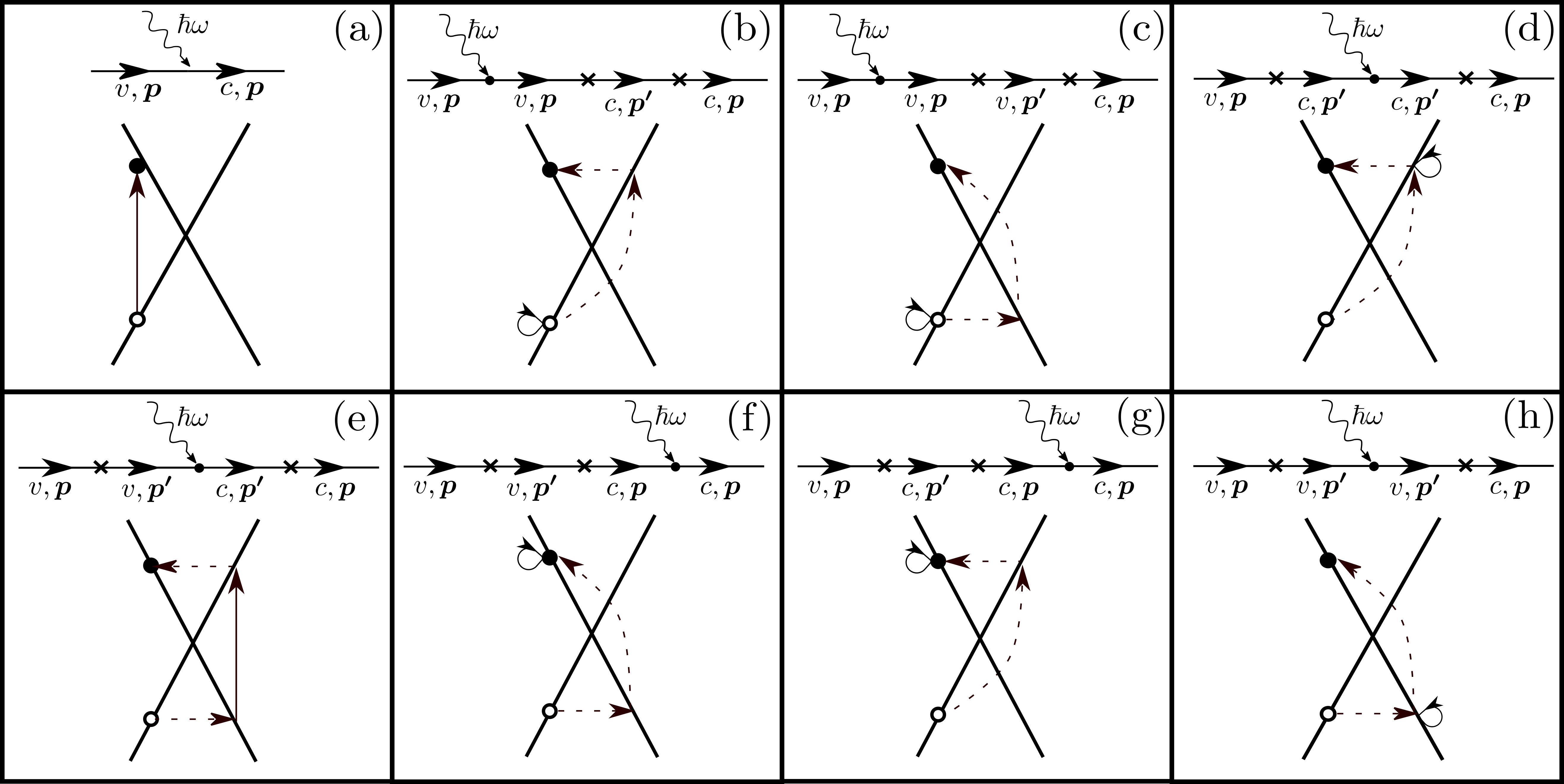} 
	\caption{First-order (a) and third-order (b)-(h) contributions to the matrix element of the transition $v\bm p \rightarrow c\bm p$.  The solid lines shows optical transitions with photon absorption and dashed lines represent scattering. }
	\label{fig:1p3}
\end{figure}

There is also a correction of the 3rd order to the direct optical transition matrix element. With the first-order term~\eqref{Vcv} the total matrix element is given by
\begin{equation}
M_{cv}^{(1+3)}(\bm p)= V_{cv}(\bm p) +\sum_{n,m} {\mathcal H'_{c\bm p,n} \mathcal H'_{n,m} \mathcal H'_{m,v\bm p}\over (E_i-E_n + i0)(E_i-E_m + i0)},
\end{equation}
where the perturbation operator $\mathcal H'= V+U$. 
The corresponding processes are depicted in  Fig.~\ref{fig:1p3}.

In the case of the electron-hole symmetry $\epsilon(p) = 0$, one has $\bm v_{c\bm p}=-\bm v_{v\bm p} = v_0 \bm p/p$. 
%The energy conservation in one of the intermediate transitions discussed above that leads to $p = p'$. 
Using relations~\eqref{eq:PU_rel} and the energy conservation in one of the intermediate transitions discussed above that leads to $p = p'$, we obtain 
${G_{\bm p' \bm p}=-G_{\bm p \bm p'} }$.
%$W_{c\bm p',v\bm p} = -W_{c\bm p,v\bm p'}$. 
As a result, it follows from Eq.~\eqref{eq:kin:ur} that $\Delta f_{c \bm p} = -\Delta f_{v \bm p}$, and the current $\bm j^{(2+2)}_{\text{ball}}$ due to $2+2$ processes is zero. 
%under the conditions of the electron-hole symmetry. 
This is also true for the current $\bm j^{(1+3)}_{\text{ball}}$ caused by the $1+3$ transitions: the pairs of depicted in Fig.~\ref{fig:1p3} processes (b) and (f), (c) and (g), and (d) and (h) cancel each other, and the process (e) does not contribute at all~\cite{note:suppl}.
% (see the SI for calculation details).
%See Supplemental Material at ... for details of derivations, which includes
%Ref.~\cite{ivchenko2012superlattices}.

With account for the electron-hole asymmetry, ${\epsilon(p)\neq 0}$, the current $\bm j^{(2+2)}_{\text{ball}}$ arises in the first order in $\dd \epsilon(p)/ \dd p$ due to the difference in the velocities, relaxation times  and densities of states in the $c$ and $v$ bands in the processes (a) and (b) in Fig.~\ref{fig:2p2}. The contribution $\bm j^{(1+3)}_{\text{ball}}$ is nonzero due to violation of cancellation of the processes (b) and (f) in Fig.~\ref{fig:1p3} caused by the corrections to the velocities and densities of states.
We considered two elastic scattering potentials: the short-range with $U_{\bm p\bm p'}=const$ and Coulomb impurities with $U_{\bm p\bm p'}\propto 1/\abs{\bm p-\bm p'}$. 
First we present the results for the linear in intensity regime where $\Delta f_{c,v}=0$ in the right-hand side of 
%$f_{c\bm p}=0$, $f_{v\bm p}=1$ 
the kinetic Eq.~\eqref{eq:kin:ur},
and the LPGE current is given by the first term of Eq.~\eqref{phenom_C3v}.
Despite a relation between the contributions $j^{(2+2)}_{\text{ball}}$ and $j^{(1+3)}_{\text{ball}}$ depends on the type of the scattering potential, their sum is the same in both cases and reads
\begin{equation}
\label{eq:chi_1_ball}
\chi_1^\text{ball}(\omega) = {3 \lambda e^3 \over 16 v_0^3} {\dd \epsilon \over \dd p}\biggr|_{p=\hbar\omega/(2v_0)}.
\end{equation}

\mysection{Shift current}
The total LPGE current is a sum of the ballistic and shift contributions which have generally the same order of magnitude~\cite{Belinicher_Ivchenko_Sturman_1982}.
The shift current is caused by accumulation of elementary shifts of electron wavepackets in the course of the optical transitions. At direct optical transitions $v \to c$ considered here the elementary shift is given by $\bm R_{cv}(\bm p)=-\bm\nabla_{\bm p}\text{arg}\qty(V_{cv})+\bm \Omega_c(\bm p) - \bm \Omega_v(\bm p)$ with $V_{cv}(\bm p)$ being the matrix element~\eqref{Vcv}, and $\bm \Omega_{c,v}$ are the Berry curvatures of the conduction and valence bands $\bm \Omega_{n}=i\hbar\bra{\psi_n}\bm \nabla_{\bm p}\ket{\psi_n}$ with $\psi_{c,v}$ being the envelopes~\eqref{psi_cv}.
%$i\hbar\expval{\bm \nabla_{\bm p}}{\psi_n}$
Calculation shows that 
%for arbitrary orientation of the light polarization plane 
the shift vector reads
\begin{equation}
\bm R_{cv}(\bm p) = {6\lambda p \qty[\sin(\theta_{\bm p}+\alpha)\hat{\bm x} - \cos(\theta_{\bm p}-\alpha)\hat{\bm y}] \over v_0\sin(\alpha-\theta_{\bm p})}.
\end{equation}
%where $\alpha$ is an angle between the polarization plane and the mirror-reflection plane $(zx)$.

The shift LPGE current is given by
\begin{equation}
\label{eq:shift_current}
	\bm j_{\text{shift}} = e \sum_{\bm p} \bm R_{cv}(\bm p)  G_{\bm p}%W_{cv}(\bm p) 
	\qty(f_{c\bm p}-f_{v\bm p}),
\end{equation}
where $G_{\bm p}=(2\pi/\hbar)\abs{V_{cv}(\bm p)}^2\delta(\varepsilon_{c\bm p}-\varepsilon_{v\bm p}-\hbar\omega)$.
In the linear in intensity regime where $f_{c\bm p}=0$, ${f_{v\bm p}=1}$, the LPGE shift current equals to the first term of Eq.~\eqref{phenom_C3v} with
\begin{equation}
\label{eq:chi_1_shift}
\chi_1^\text{shift} = -{3 \lambda e^3 \over 8 v_0^2}.
\end{equation}
This expression corrects the shift current contribution calculated in Ref.~\cite{Kim_Shift_2017}, see~\cite{note:suppl}.

\mysection{Nonlinear in intensity regime}
The nonlinearity in the photocurrent comes from saturation of the optical transitions due to changes in the electron occupations in the conduction- and valence bands~\cite{belinicher1980photogalvanic,Nonlin_PGE1,pssb_2019,Matsyshyn2021,Nonlin_our_PRB_2022}. 
%We find the corrections to the occupations $\Delta f_{c,v}$ in the first order in the light intensity accounting for them in right-hand side of the kinetic Eq.~\eqref{eq:kin:ur}.
%\NL{\sout{The collision integrals account for both the elastic momentum- and inelastic energy relaxation
%in the relaxation-time approximations, see SI for details. }}
Considering the optical transition saturation effect on the photocurrent it is essential to take into account inelastic decay from the photoexcited states~\cite{Mishchenko2009,pssb_2019,Nonlin_our_PRB_2022}. Following Ref.~\cite{Nonlin_our_PRB_2022} we take the collision integral in the relaxation-time approximation introducing the energy relaxation time $\tau_\varepsilon$~\cite{note:suppl}.
%\NL{If one consider the optical transition saturation effect on the photocurrent it is essential to take into account the decay from the states with the energy of the generation~\cite{Mishchenko2009,pssb_2019,Nonlin_our_PRB_2022}. Following the work~\cite{Nonlin_our_PRB_2022} we introduce it in the collision integral (see SI) using relaxation time $\tau_\varepsilon$.
As a result, the transport relaxation rate reads $1/\tau_1 = 1/\tau_1^*+1/\tau_\varepsilon$ with $\tau_1^*$ being  the transport relaxation time caused by elastic scattering, 
and the linear in the intensity ballistic current has the form
%In this case the current relaxation rate $1/\tau_1 = 1/\tau_1^*+1/\tau_\varepsilon$ and the linear in the intensity ballistic current in this case needs to be multiplied by the factor $\tau_1/\tau_1^*$
\begin{equation}
	\tilde{\chi}_1^\text{ball}(\omega) = \frac{\tau_1}{\tau_1^*} \chi_1^\text{ball}(\omega).
\end{equation}
%}
We find the corrections to the occupations $\Delta f_{c,v}$ in the first order in the light intensity accounting for them in right-hand side of the kinetic Eq.~\eqref{eq:kin:ur}.
Calculating the ballistic current we account for the non-Born contributions to the transition matrix element $M_{cv}$ as in the linear in intensity regime. As a result,
%by Eq.~\eqref{eq:ball_general},
we obtain Eq.~\eqref{phenom_C3v} with
\begin{align}
\label{eq:d_chi}
&\Delta \chi_1^{\rm ball} = -\tilde{\chi}_1^{\rm ball}(\omega) F^2\qty(1+{3 \tau_1 \over 4 \tau_\varepsilon} -{11 \tau_2 \over 24 \tau_\varepsilon}), \nonumber
\\
&\chi_2^{\rm ball}E^2 = \tilde{\chi}_1^{\rm ball}(\omega) F^2 {12\tau_1+\tau_2\over 24 \tau_\varepsilon}.
%\qty({\tau_1\over 2 \tau_\varepsilon} +{\tau_2\over 24 \tau_\varepsilon}).
\end{align}
Here $F=2eEv_0\tau_\varepsilon/(\hbar\omega)$ and 
%\sout{$\tau_\varepsilon$ is the energy relaxation time, and $1/\tau_{1,2}=1/\tau_\varepsilon + 1/\tau_{1,2}^*$} 
$1/\tau_{2}=1/\tau_\varepsilon + 1/\tau_{2}^*$  with $\tau_{2}^*$ being the elastic relaxation time 
%describing relaxation 
of the second angular harmonics of the distribution function. In Eqs.~\eqref{eq:d_chi} we assumed that elastic scattering occurs at short-range impurities.
Finding $\Delta f_{c,v}$ in the opposite limit $F \gg 1$ we obtain the asymptotic~\cite{note:suppl}
\begin{equation}
\label{asympt_ball}
	j_x^{\rm ball}(F \to \infty) = -\tilde{\chi}_1^{\rm ball} E^2  \frac{\ln F}{6 \pi F}(\cos2\alpha-\cos4\alpha).
%	j_x^{\rm ball}(I \to \infty) = \tilde{\chi}_1^{\rm ball} E^2  \times \frac{10-\ln(64)-3\ln(F)}{18 F \pi}(\cos(2\alpha)-\cos(4\alpha))
\end{equation}

For calculation of the intensity dependence of the shift photocurrent, one needs to account for 
%the first term in the lhs of the kinetic Eq.~\eqref{eq:Kin_eq} 
direct optical transitions with the matrix element $M_{cv}=V_{cv}(\bm p)$ only, and ignore both the warping and the electron-hole asymmetry. The result is that only the first contribution to Eq.~\eqref{phenom_C3v} is present:
\begin{equation}
\Delta \chi_1^{\rm shift} = \chi_1^{\rm shift}{F^2\over 2}\qty(1 + {\tau_2\over 2\tau_\varepsilon}), 
\quad \chi_2^{\rm shift} =0.
\end{equation}
The shift LPGE current can be calculated in all orders in the light intensity. We obtain that the shift contribution has the same dependence on the dimensionless electric field amplitude 
%$\mathcal E=eEv_0\tau_\varepsilon/(\hbar\omega)$ 
as the absorbance at linear polarization $\eta_{\rm lin}(E)$: 
\begin{equation}
\chi_1^{\rm shift}(E) =  -{3 \lambda e^3 \over 8 v_0^2} \times {\eta_{\rm lin}(E) \over \eta_0},
\end{equation}
where $\eta_0=\pi e^2/(4\hbar c)$ is the low-intensity absorbance. The dependence of the absorbance on the light amplitude was analyzed in detail in Ref.~\cite{Nonlin_our_PRB_2022} where it was shown that it is strongly sensitive to the relation between the energy and momentum relaxation rates. In the particular case of fast energy relaxation ($\tau_\varepsilon \ll \tau_{1,2}^*$) we have~\cite{note:suppl}
\begin{equation}
\label{eq:shift_nl}
\chi_1^{\rm shift}(E) =  -{3 \lambda e^3 \over 2\pi v_0^2} \times {\text{K}(m)-(1+F^2)\text E(m)\over F^2 \sqrt{1+F^2}}.
%\chi_1^{\rm shift}(E) =  {3 \lambda e^3 \over 2\pi v_0^2} \times {\text{K}(m)-(1+\mathcal E^2)\text E(m)\over \mathcal E^2 \sqrt{1+\mathcal E^2}}.
\end{equation}
Here 
%\NL{\sout{$F=2eEv_0\tau_\varepsilon/(\hbar\omega)$,}} 
%$m=\mathcal E^2/(1+\mathcal E^2)$, 
$m=F^2/(1+F^2)$, and 
$\text E(m)[\text K(m)] = {\int_0^{\pi/2} \dd \theta (1 - m \sin^2\theta)^{\pm 1/2} }$ are the complete elliptic integrals.
The high-intensity asymptotic at $F \gg 1$ reads 
%(see SI for details)
\begin{equation}
\label{asympt_shift}
j_x^{\rm shift}(F \to \infty) = \chi_1^{\rm shift} E^2   \frac{4}{\pi F}\cos2\alpha.
\end{equation}

\mysection{Discussion}
The obtained expressions for the ballistic and shift contributions to the LPGE current demonstrate a possibility to distinguish them in experiments. 
Comparing Eqs.~\eqref{eq:chi_1_ball} and~\eqref{eq:chi_1_shift} we see that, by contrast to the ballistic contribution, the shift current is nonzero in electron-hole symmetric systems. At the same time, the frequency dependence is present in $\chi_1^\text{ball}$ due to the correction to the electron velocity $\dd \epsilon/\dd p$ while $\chi_1^\text{shift}$ is independent of frequency. For example, if $\epsilon(p)$ is parabolic in momentum, the ballistic contribution linearly raises with frequency.
This allows for their separation 
%by study of the LPGE current dependence on light frequency 
already in the linear in the light intensity regime.

\begin{figure}[htp]
	\centering
	\includegraphics[width=1\linewidth]{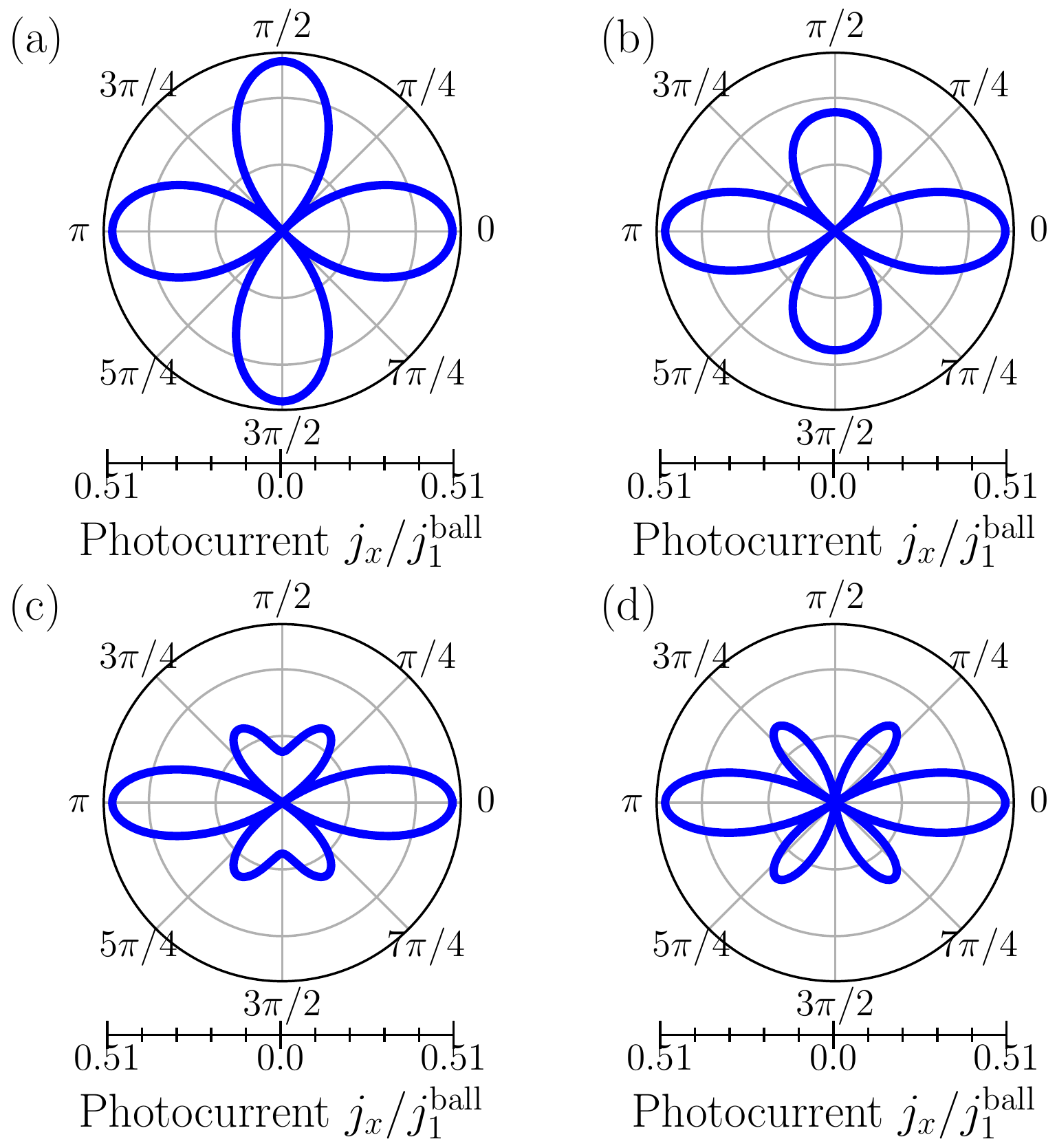} 	
	\caption{Dependence of the ballistic LPGE current at $F = 0.7$ on the polarization plane orientation for different relations between elastic and inelastic scattering: (a)~$\tau_1/\tau_\varepsilon\to 0$, (b)~$\tau_1/\tau_\varepsilon = 0.3$, (c)~$\tau_1/\tau_\varepsilon = 0.7$, (d)~$\tau_1/\tau_\varepsilon = 1$. 
	%The value of the field $F = 0.7$.
	}
	\label{fig:angular_ball}
\end{figure}

%More details of 
Deeper insight into 
the ballistic photocurrent could be seen in the nonlinear in intensity regime. 
It is shown in Fig.~\ref{fig:angular_ball} that the dependence on the polarization plane orientation is 
controlled by the relation between elastic and inelastic relaxation times. 
At slow energy relaxation, $\tau_1^*,\tau_2^* \ll  \tau_\varepsilon$, 
the second angular harmonics dominates over the fourth one, Fig.~\ref{fig:angular_ball}(a): $\chi_2^{\text{ball}}E^2\ll \tilde{\chi}_1^{\text{ball}} + \Delta \chi_1^{\text{ball}}$.
This happens because for long energy relaxation $\tau_\varepsilon$ the saturation of the optical transition occurs and the high angular harmonics of the distribution function are not generated. In the opposite limit $\tau_1^*,\tau_2^* \gg  \tau_\varepsilon$ 
the distribution function keeps its anisotropy for longer time which results in a strong
%higher distribution harmonics are generated, 
%for the time of saturation, 
%and we obtain the stronger 
fourth harmonics of the photocurrent, see Fig.~\ref{fig:angular_ball}(b)-(d).

The dependence on the polarization plane orientation for the shift current differs strongly. Even in the nonlinear regime, the shift current is still described by the first term in Eq.~\eqref{phenom_C3v} with $\chi_1^{\rm shift}(E)=\chi_1^{\rm shift}+\Delta \chi_1^{\rm shift}(E)$. This means that the contribution to the total LPGE current described by $\chi_2$ 
has a pure ballistic nature. Moreover, at very high light intensities the ballistic current dominates over the shift one. Comparing their asymptotes~\eqref{asympt_ball} and~\eqref{asympt_shift} at $F \gg 1$ we conclude that, due to an additional logarithmic contribution, the ballistic contribution is stronger. As a result, at so high light intensities, the amplitudes of the second and fourth polarization harmonics have the same amplitude.
%\NL{This is in contrast to the shift current which in the nonlinear regime is still described by the first term in Eq.~\eqref{phenom_C3v} with the electric field dependent $\chi_1^{\rm shift}(E)=\chi_1^{\rm shift}+\Delta \chi_1^{\rm shift}(E)$. This means that the LPGE current described by $\chi_2$ 
%has a pure ballistic nature.}

\mysection{Summary} The theory of the LPGE in topological insulators is developed with account for all microscopic contributions to the photocurrent. 
We demonstrated that the ballistic contribution, formed in the course of elastic scattering, is independent of the concentration of scatterers and is parametrically comparable to the shift contribution. At low light intensity, $\bm j_{\rm ball}(\omega)$ is frequency dependent in contrast to $\bm j_{\rm shift}$. In the nonlinear in the light intensity regime, the ballistic photocurrent has contributions varying differently with the polarization plane orientation which allows for their direct measurements. At very high intensity, the ballistic contribution dominates.

\mysection{Acknowledgements} We thank E. L. Ivchenko for helpful discussions. The work was supported by the Russian Science Foundation (Project No. 20-12-00147) and the Foundation for the
Advancement of Theoretical Physics and Mathematics ``BASIS''.

%----------------------------------
\renewcommand{\i}{\ifr}
\let\oldaddcontentsline\addcontentsline% Store \addcontentsline
\renewcommand{\addcontentsline}[3]{}% Make \addcontentsline a no-op

\bibliography{Trig_TI.bbl}

%\bibliography{citations}% Produces the bibliography via BibTeX.

\end{document}

\title{Linear photogalvanic effect in surface states of topological insulators}

\author{N.~V.~Leppenen}
\email{leppenen@mail.ioffe.ru}
%\affiliation{Ioffe Institute, 194021 St. Petersburg, Russia}
%
\author{L.~E.~Golub}
\affiliation{Ioffe Institute, 194021 St. Petersburg, Russia}

%\appendix

%%%%%%%%%%%%%%%%%%%%               SOM               %%%%%%%%%%%%%%%%%%%%%%%%%%%%%%%%%
\onecolumngrid
\vspace{\columnsep}
\begin{center}
\makeatletter
{\large\bf{Supplementary Information\\``Linear photogalvanic effect in surface states of topological insulators''}}
\makeatother
\end{center}
\vspace{\columnsep}

The Supplementary Material includes the following topics:

\hypersetup{linktoc=page}
\tableofcontents
\vspace{\columnsep}

\counterwithin{figure}{section}
\renewcommand{\thesection}{S\arabic{section}}
\renewcommand{\section}[1]{\oldsec{#1}}
\renewcommand{\thepage}{S\arabic{page}}
\renewcommand{\theequation}{S\arabic{equation}}
\renewcommand{\thefigure}{S\arabic{figure}}
\renewcommand{\bibnumfmt}[1]{[S#1]}
\renewcommand{\citenumfont}[1]{S#1}

\setcounter{page}{1}
\setcounter{section}{0}
\setcounter{equation}{0}
\setcounter{figure}{0}

\section{Surface states in the presence of warping and electron-hole asymmetry}
%\label{sec:Model}

Conduction and valence surface bands in topological
insulators are described by the linear in momentum effective Hamiltonian

\begin{equation}\label{eq:Ham}
	{\cal H}_0(\bm p) =  \epsilon(p)+v_0 \qty[\bm \sigma \times \bm p]_z,
\end{equation}
where $z$ is a direction normal to the surface, $\sigma_{x,y}$ are Pauli matrices, and $v_0$ is the surface state's velocity. Due to the C$_{3v}$ symmetry of the system, there is a correction to the Hamiltonian:
\begin{equation}
	{\cal H}_w(\bm p) = \lambda \sigma_z \frac{p_+^3-p_-^3}{2i}.
\end{equation}
With account for the trigonal term in the first order, the wavefunctions have the form 
\begin{equation}
\label{eq:S:psi_c_v}
	\ket{c,\bm p} = \frac{1}{\sqrt{2}}\mqty[ie^{-i\theta_{\bm p}/2} \qty(1+\Lambda_{\bm p}) \\ e^{i\theta_{\bm p}/2}(1-\Lambda_{\bm p})],\qquad \ket{v,\bm p} = \frac{1}{\sqrt{2}}\mqty[-ie^{-i\theta_{\bm p}/2} \qty(1-\Lambda_{\bm p}) \\ e^{i\theta_{\bm p}/2}(1+\Lambda_{\bm p})],
\end{equation}
where 
\begin{equation}
	\Lambda_{\bm p} = \frac{\lambda p^2 \sin 3\theta}{2v_0}.
\end{equation}
For the arbitrary form of the the $\epsilon(p)$ the energy spectrum in the first order of $\lambda$ is given by 
\begin{equation}
	\varepsilon_{c,v}(\bm p) = \epsilon(p)\pm v_0 p+{\cal O}(\lambda^2).
\end{equation}

\subsection{Optical transition matrix elements}
The electron-photon interaction Hamiltonian is given by 
\begin{equation}
	V = \frac{i e}{\omega}\bm E \cdot \grad_{\bm p}[\mathcal{H}_0(\bm p)+\mathcal{H}_w(\bm p)]
\end{equation}
Calculating the matrix element of a direct optical transition from the valence to the conduction surface band $V_{cv} = \mel{c}{V}{v}$ we obtain up to the first
order in $\lambda$
\begin{equation}\label{Seq:mat_el_1}
V_{cv}(\bm p)  = -\frac{e E v_0}{\omega}\qty{\sin(\theta_{\bm p}-\alpha)+\frac{i\lambda p^2}{2 v_0} \qty[5\sin(2\theta_{\bm p}+\alpha)-\sin(4\theta_{\bm p}-\alpha)]}+{\cal O}(\lambda^2),
\end{equation}
where we assumed
\begin{equation}
	\bm E = E \qty(\cos\alpha \bm e_x + \sin\alpha \bm e_y).
\end{equation}
The indirect optical transitions have no first-order in $\lambda$ corrections and are given by 
\begin{subequations}\label{eq:v_intra}
\begin{equation}
	V_{vv}(\bm p)   = \frac{i e}{\omega}\bm E \cdot \bm v_{v, \bm p}= -\frac{ieE}{\omega} \qty(v_0 -\dv{\epsilon(p)}{p})\cos(\theta_{\bm p}-\alpha)+{\cal O}(\lambda^2),
\end{equation} 
\begin{equation}
	V_{cc}(\bm p)  = \frac{i e}{\omega}\bm E \cdot \bm v_{c, \bm p} = \frac{ieE}{\omega} \qty(v_0 +\dv{\epsilon(p)}{p})\cos(\theta_{\bm p}-\alpha)+{\cal O}(\lambda^2),
\end{equation}
\end{subequations}
where $\bm v_{(c,v),\bm p}=\bm \nabla_{\bm p} \varepsilon_{(c,v),\bm p} = (v_0\pm \dd \epsilon(p)/\dd p){\bm p\over p}$ are the electron velocities in the bands.

\subsection{Impurity scattering matrix elements}
We consider scalar impurity scattering with matrix elements
\begin{equation}
	U_{n',\bm p';n,\bm p} = U_{\bm p'\bm p}\bra{n'\bm p'}\ket{n,\bm p},
\end{equation} 
where $ U_{\bm p'\bm p} = U_0$ for the short-range impurity potential, and $U_{\bm p'\bm p} = \frac{2\pi e^2}{\varkappa \abs{\bm p-\bm p'}}$ for the Coulomb potential ($\varkappa$ is the dielectric constant).
 For the intraband scattering $c,\bm p \to c,\bm p'$ the matrix element has the form
\begin{equation}\label{eq:Ucc}
	U_{c,\bm p';c,\bm p} =U_{\bm p'\bm p} \left\{\cos\qty(\frac{\theta_{\bm p'}-\theta_{\bm p}}{2})+i\frac{\lambda p^2}{v_0}\sin\qty(\theta_{\bm p'}-\theta_{\bm p}\over 2) \sin\qty[3(\theta_{\bm p'}+\theta_{\bm p})\over 2]\cos\qty[3(\theta_{\bm p'}-\theta_{\bm p})\over 2]\right\}.
\end{equation}
The interband scattering $v,\bm p \rightarrow c,\bm p'$ does not obey the energy conservation but will be playing important role in virtual transitions for the ballistic current
\begin{equation}\label{eq:Ucv}
 	U_{c,\bm p';v,\bm p} = U_{\bm p'\bm p} \left\{i\sin\qty(\frac{\theta_{\bm p}-\theta_{\bm p'}}{2})- \frac{\lambda p^2}{v_0}\cos\qty(\theta_{\bm p'}-\theta_{\bm p}\over 2)\sin\qty[3(\theta_{\bm p'}-\theta_{\bm p})\over 2]\cos\qty[3(\theta_{\bm p'}+\theta_{\bm p})\over 2]\right\}.
 \end{equation}
 
The relation between the matrix elements of intraband scattering in the valence band and interband $c,\bm p \rightarrow v,\bm p'$ scattering for the system described by the Hamiltonian $2 \times 2$ could be obtained from the time-reversal symmetry.
The time reversal operator ${\cal T}$ connects the states with the same energy. Let us choose the phase of ${\cal T}$ so that
\begin{equation}
	\ket{c,-\bm p} = {\cal T}\ket{c,\bm p}.
\end{equation}

Since ${\cal T}$ is anti-unitary, it is straightforward to show
\begin{equation}
	\bra{c,\bm p}\ket{c,-\bm p}  = \qty({\cal T}\ket{c,-\bm p})^\dagger {\cal T}\ket{c,\bm p} = -\bra{c,\bm p} \ket{c,-\bm p} =0,
\end{equation}
which means an absence of the backscattering.
On the other hand, since $\ket{c,\bm p}$ and $\ket{v,\bm p}$ are the eigenvectors of a $2 \times 2$ Hamiltonian,
% the Hamiltonian ${\cal H}(p)$, 
they form the full basis and 
\begin{equation}
	{\cal T}\ket{c,\bm p} = \ket{c,-\bm p} =  \bra{c,\bm p}\ket{c,-\bm p}\ket{c,\bm p}+ \bra{v,\bm p}\ket{c,-\bm p}\ket{v,\bm p} = \bra{v,\bm p}\ket{c,-\bm p}\ket{v,\bm p}.
\end{equation}

Due to normalization, we can write $ \bra{v,\bm p}\ket{c,-\bm p} = e^{i\gamma}$ with a real $\gamma$. Thus we proof that for %the Hamiltonian $2 \times 2$ 
any $2 \times 2$ Hamiltonian
the following relation holds
\begin{equation}
	-\ket{c,\bm p} = e^{-i\gamma}{\cal T}\ket{v,\bm p} \equiv \Theta \ket{v,\bm p},
\end{equation}
where $\Theta$ is also antiunitary
\begin{equation}
\Theta^2 = -1,\qquad \Theta \ket{c,\bm p} = \ket{v,\bm p},\qquad \Theta \ket{v,\bm p} = -\ket{c,\bm p},
\end{equation} 
and has the property~\cite{shen2012topological}
\begin{equation}\label{eq:TUT}
	\mel{\Theta \alpha}{U}{\Theta \beta} = \mel{\beta}{U}{\alpha},
\end{equation}
for any $U$ that is time reversal invariant. Using these relations one can show
\begin{subequations}\label{eq:proof1}
\begin{equation}
	\mel{c,\bm p'}{U}{v,\bm p} = \qty(\Theta \ket{v,\bm p})^\dagger U\Theta\ket{c,\bm p'}= - \mel{v,\bm p}{U}{c,\bm p'}.
\end{equation}
\begin{equation}
	\mel{c,\bm p'}{U}{c,\bm p} =   \qty(\Theta \ket{c,\bm p})^\dagger U\Theta\ket{c,\bm p'} = \mel{v,\bm p'}{U}{v,\bm p}.
\end{equation}
\end{subequations}
Thus, 
%it follows that
we obtained for any order of the electron-hole asymmetry and warping
\begin{equation}\label{eq:U_rel}
	U_{v,\bm p';v,\bm p} = U_{c,\bm p; c,\bm p'}, \qquad U_{c,\bm p';v,\bm p} = -U_{c,\bm p; v,\bm p'}.
\end{equation}
These 
%Note that this 
relations will play an important role in the calculation of the ballistic current.
% and are valid for any order of the electron-hole asymmetry and warping.

\subsection{Relaxation times}

%In the work we will 
We
consider both short-range and Coulomb potentials for the elastic scattering. Due to the electron-hole asymmetry introduced in Eq.~\eqref{eq:Ham} by the term $\epsilon(p)$ the relaxation times are different in different bands. Here we neglect the warping in the squared absolute values of the scattering matrix elements since it contributes only in the second order of $\lambda$, see Eq.~\eqref{eq:Ucc}.

The relaxation time in this case is calculated using the scattering rate $W^i_{\bm p'\bm p}$ and for the relaxation of the $n$-th Fourier-harmonics of the distribution function in the band $i = c,v$ is given by 
\begin{equation}
	\frac{1}{\tau_i^{(n)}} = \sum_{\bm p'}W^i_{\bm p'\bm p}(1-\cos n \theta_{\bm p,\bm p'}),
\end{equation}
where $W^i_{\bm p'\bm p}$ is given by the Fermi Golden rule 
\begin{equation}\label{eq:FGR}
	W^i_{\bm p',\bm p} = 2\pi \abs{U_{i,\bm p';i,\bm p}}^2(f_{i,\bm p}-f_{i,\bm p'})\delta(\varepsilon_{c,\bm p'}-\varepsilon_{c,\bm p}),
\end{equation}
where $U_{i,\bm p';i,\bm p}$ are given by Eqs.~\eqref{eq:Ucc} and~\eqref{eq:U_rel}. Hereafter we set $\hbar \equiv 1$.

For the short-range potential
\begin{equation}\label{eq:SR_tau}
	\text{Short-range:} \qquad \frac{1}{\tau^{(1)}_{(c,v)}(p)} = {\pi\over 2} N_{\text{imp}} \abs{U_0}^2 {\cal D}_{(c,v)}(p),\quad \frac{1}{\tau^{(n\geq 2)}_{(c,v)}(p)} = \frac{2}{\tau^{(1)}_{(c,v)}(p)},
\end{equation}
and for the Coulomb potential 
\begin{equation}\label{eq:Coul_tau}
		\text{Coulomb:} \qquad \frac{1}{\tau^{(1)}_{(c,v)}(p)} = \frac{2\pi^3 N_{\text{imp}}e^4{\cal D}_{(c,v)}(p)}{\varkappa^2 p^2},\quad \tau^{(2)}_{(c,v)}(p) =\tau^{(1)}_{(c,v)}(p)/3,
\end{equation}
where $N_{\text{imp}}$ is the number of the impurities and 
\begin{equation}
	{\cal D}_{(c,v)}(p) =  \frac{p}{2\pi \abs{v_0\pm \dd{\epsilon(p)}/\dd{p}}} \approx \frac{p}{2\pi v_0}\qty(1\mp \frac{1}{v_0}\dv{\epsilon(p)}{p})
\end{equation}
are densities of states in the bands.

The calculation of the nonlinear in the light intensity current in the Sec.~\ref{sec:NonlinBall} requires accounting for inelastic scattering~\cite{Nonlin_our_PRB_2022}. We denote the corresponding relaxation time $\tau_\varepsilon$. In this case the total relaxation rates of the first and second angular harmonics are given by 
\begin{equation}
	\frac{1}{\tau_{1,2}} = \frac{1}{\tau_\varepsilon}+\frac{1}{\tau_{1,2}^*},
\end{equation}
where $\tau_{1,2}^*$ is given by Eqs.~\eqref{eq:SR_tau}-\eqref{eq:Coul_tau} after the replacement ${\cal D}_{(c,v)}(p) \rightarrow {\cal D}_0$ where 
\begin{equation}
	{\cal D}_0 = \frac{p}{2\pi v_0}
\end{equation}
is the independent of the electron-hole asymmetry density of states.

\section{Ballistic contribution to the current}
The ballistic current is described by the expression
\begin{equation}\label{Seq:b_cur}
	\bm j_{\text{ball}} = e \sum_{\bm p} \qty(\bm v_{c\bm p} f_{c\bm p}+ \bm v_{v\bm p}f_{v\bm p}) ,
\end{equation}
where in the first order of the light intensity considering elastic scattering only, the correction to the distribution function $f_{(c,v),\bm p} = f_0\qty(\varepsilon_{(c,v),\bm p})+\Delta f_{(c,v),\bm p}$ is given by 
\begin{equation}
\label{eq:Delta_f}
	\Delta f_{(c,v),\bm p} = \pm \tau^{(1)}_{(c,v)} \sum_{\bm p'}G_{\bm p, \bm p'}.
\end{equation}
Here $G_{\bm p \bm p'}
=2\pi\abs{M_{c\bm p, v\bm p'}}^2 \delta(\varepsilon_{c\bm p}-\varepsilon_{v\bm p'}-\omega)$ is the generation rate.

The ballistic contribution calculation should include the interaction with phonons  or impurities~\cite{belinicher1980photogalvanic,andrianov1981linear,
ivchenko2012superlattices}. We consider elastic impurity  scattering and include all possible transitions. The contribution in the first-order in $\lambda$  comes from the terms in $\abs{M}^2$ associated with the interference of two compound matrix elements of the second order, or of the first and third orders in the perturbation $\mathcal H'=V+U$~\cite{ivchenko2012superlattices}. 
\subsection{Interference of the second order processes}
\begin{figure}[htp]
	\centering
	\includegraphics[scale=0.4]{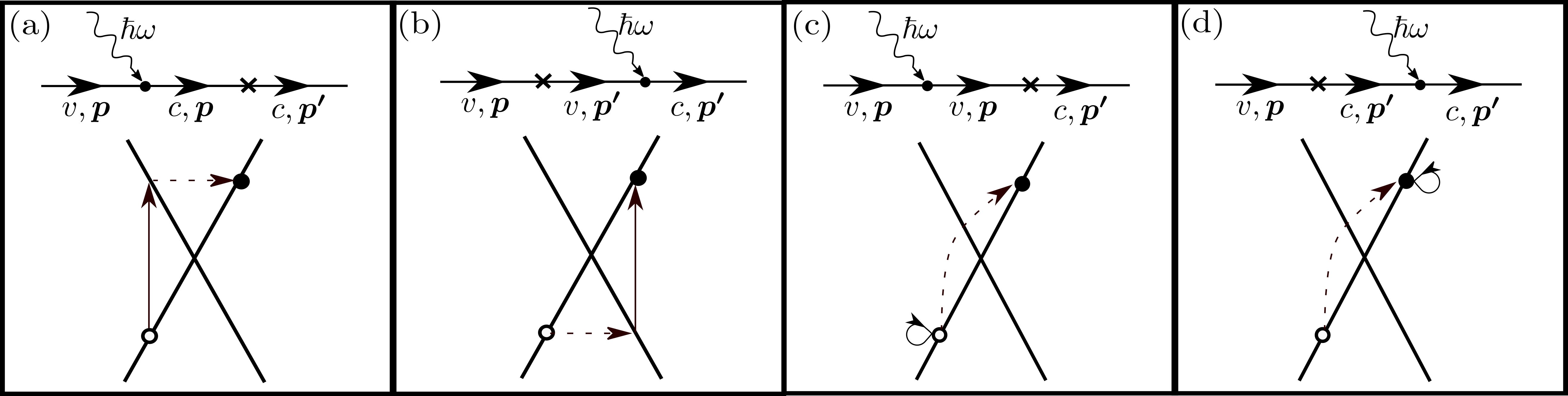} 
	\caption{Second order contributions to the matrix element of the transition $v,\bm p \rightarrow c,\bm p'$. The solid lines show interband (a,b) or intraband (c,d) optical transitions and dashed lines represent intraband (a,b) and interband (c,d) scattering.}
	\label{fig:S2p2}
\end{figure}

For calculation of the ballistic contribution from transitions shown in Fig.~\ref{fig:S2p2} we rewrite Eqs.~\eqref{Seq:b_cur},~\eqref{eq:Delta_f} as
\begin{equation}\label{eq:ball_2p2}
	\bm j_{(2+2)}^{\text{ball}} = e \sum_{\bm p,\bm p'} G_{\bm p'\bm p} \qty(\bm v_{c,\bm p'}\tau^{(1)}_{c,\bm p'}-\bm v_{v,\bm p}\tau^{(1)}_{v,\bm p}).
\end{equation}

The generation rate $G_{\bm p'\bm p}$ is caused by interference of the processes shown in Fig.~\ref{fig:S2p2}. The corresponding matrix elements read
\begin{subequations}
\begin{equation}\label{eq:2a}
	M^{(a)}_{c,\bm p';v,\bm p} = \frac{V_{cv}(\bm p)U_{c,\bm p';c,\bm p}}{\varepsilon_{v,\bm p}+\hbar \omega-\varepsilon_{c,\bm p}+i0} =  \frac{V_{cv}(\bm p)U_{c,\bm p';c,\bm p}}{\varepsilon_{c,\bm p'}-\varepsilon_{c,\bm p}+i0},
\end{equation}
\begin{equation}\label{eq:2b}
	M^{(b)}_{c,\bm p';v,\bm p} = \frac{U_{v,\bm p';v,\bm p}V_{cv}(\bm p')}{\varepsilon_{v,\bm p}-\varepsilon_{v,\bm p'}+i0}, 
\end{equation}
\begin{equation}\label{eq:2c}
	M^{(c)}_{c,\bm p';v,\bm p} = \frac{V_{vv}(\bm p)U_{c,\bm p';v,\bm p}}{\varepsilon_{v,\bm p}+\omega-\varepsilon_{v, \bm p}+i0} = \frac{V_{vv}(\bm p)U_{c,\bm p';v,\bm p}}{\omega},
\end{equation}
\begin{equation}\label{eq:2d}
	M^{(d)}_{c,\bm p';v,\bm p} = \frac{U_{c,\bm p';v, \bm p}V_{cc}(\bm k')}{\varepsilon_{v, \bm p}+ \omega-\varepsilon_{c,\bm p'}- \omega+i0} =- \frac{U_{c,\bm p';v, \bm p}V_{cc}(\bm p')}{\omega}. 
\end{equation}
\end{subequations}
In the denominators of matrix elements~\eqref{eq:2a} and~\eqref{eq:2d} we used the energy conservation law $\varepsilon_{c,\bm p'} = \varepsilon_{v,\bm p}+\omega$ that is provided by the $\delta(\varepsilon_{c,\bm p'}-\varepsilon_{v,\bm p}-\omega)$ in $G_{\bm p',\bm p}$. 
The denominators in~\eqref{eq:2a} and~\eqref{eq:2b} could be rewritten using Sokhotski-Plemelj theorem
\begin{equation}
	\frac{1}{x-y+i\delta} = \PV\frac{1}{x-y}-i\pi \delta(x-y).
\end{equation}
%Due to the time-inversion the $W_{c,\bm p';v,\bm p}$ should include one dissipative term, those we need to take one addition delta-function that will provide the energy conservation with one of the intermediate state~\cite{ivchenko2012superlattices} in every interference term.
Due to the time-inversion symmetry arguments, only the terms containing one imaginary factor with an additional $\delta$-function contribute
 to the current. 
% will contribute terms containing one imaginary term with additional delta-function 
This means energy conservation in 
%that will provide the energy conservation with 
one of the intermediate states~\cite{ivchenko2012superlattices} . 
Therefore
in the general expression $$G_{\bm p'\bm p} \propto \abs{M^{(a)}_{c,\bm p';v,\bm p}+M^{(b)}_{c,\bm p';v,\bm p}+M^{(c)}_{c,\bm p';v,\bm p}+M^{(d)}_{c,\bm p';v,\bm p}}^2$$
%but the contribution in the first-order of $\lambda$ will arise only due to the interference terms. 
only the interference terms of processes (a) and (c), (a) and (d), (b) and (c), (b) and (d) contribute:
%Then the matrix element will include only the interference of processes (a) and (c), (a) and (d), (b) and (c), (b) and (d):
\begin{equation}
	G_{\bm p'\bm p} = 4\pi \Re[( M^{(a)}_{c,\bm p';v,\bm p}+M^{(b)}_{c,\bm p';v,\bm p})^*(M^{(c)}_{c,\bm p';v,\bm p}+M^{(d)}_{c,\bm p';v,\bm p})]\delta(\varepsilon_{c,\bm p'}-\varepsilon_{v,\bm p}-\omega),%(f_{v,\bm p}-f_{c,\bm p'})
\end{equation}
where the term proportional to the delta-function is taken in  $M^{(a)}_{c,\bm p';v,\bm p}$ and $M^{(b)}_{c,\bm p';v,\bm p}$.
After  some algebra we obtain 
\begin{multline}\label{eq:Wcv2p2}
		\Re[( M^{(a)}_{c,\bm p';v,\bm p}+M^{(b)}_{c,\bm p';v,\bm p})^*(M^{(c)}_{c,\bm p';v,\bm p}+M^{(d)}_{c,\bm p';v,\bm p})] = \\ = -{\pi\over \omega}\Im\qty{\qty[V_{cv}^*(\bm p)U^*_{c,\bm p';c,\bm p}\delta(\varepsilon_{c,\bm p'}-\varepsilon_{c,\bm p})+V_{cv}^*(\bm p')U^*_{v,\bm p';v,\bm p}\delta(\varepsilon_{v,\bm p'}-\varepsilon_{v,\bm p})]U_{c,\bm p';v,\bm p}\qty[V_{vv}(\bm p)-V_{cc}(\bm p')]}.
\end{multline}
One can see, that in the case of the electron-hole symmetry $\tau_{c,\bm p}^{(1)} =\tau_{v,\bm p}^{(1)}$ and $\bm v_{c,\bm p}=-\bm v_{v,\bm p} = v_0 \bm p/p$, so the expression in the round brackets in~\eqref{eq:ball_2p2} in symmetric under the change $\bm p \leftrightarrow \bm p'$, while the expression~\eqref{eq:Wcv2p2} is anti-symmetric due to the relations~\eqref{eq:U_rel}. Therefore in the case of the electron-hole symmetry the current~\eqref{eq:ball_2p2} is equal to zero.

In the first order of the parameter of the electron-hole asymmetry $\frac{1}{v_0}\dv{\epsilon(p)}{p}$ the current may arise due to the difference in the relaxation times, the intraband densities of states and in the velocities that included in the expression for the current~\eqref{eq:ball_2p2} and for the intraband transition matrix elements~\eqref{eq:v_intra}:
\begin{equation}
	\bm j_{(2+2)}^{\text{ball}} =\bm j_{(2+2)}^{\tau}+\bm j_{(2+2)}^{{\cal D}}+\bm j_{(2+2)}^{v}.
\end{equation}
The calculation of the contributions yields 
\begin{multline}
	\bm j_{(2+2)}^{\tau} = \frac{8\pi^2 e^2 v_0}{\omega^2}\sum_{\bm p}\expval{\Re[V_{cv}^*(\bm p)U_{c,\bm p';c,\bm p}^*U_{c,\bm p';v,\bm p}]\frac{[\bm E \cdot(\bm p+\bm p')](\bm p'-\bm p)}{p^2}} \mathcal{D}_0(p)\dv{\epsilon(p)}{p} \delta(\varepsilon_{c,\bm p}-\varepsilon_{v,\bm p}-\omega)\tau^{(1)}_{\bm p},
\end{multline}
\begin{multline}
	\bm j_{(2+2)}^{{\cal D}} = -\frac{8\pi^2 e^2 v_0}{\omega^2}\sum_{\bm p}\expval{\Re[V_{cv}^*(\bm p)U_{c,\bm p';c,\bm p}^*U_{c,\bm p';v,\bm p}]\frac{[\bm E \cdot(\bm p+\bm p')](\bm p+\bm p')}{p^2}} \mathcal{D}_0(p)\dv{\epsilon(p)}{p} \delta(\varepsilon_{c,\bm p}-\varepsilon_{v,\bm p}-\omega)\tau^{(1)}_{\bm p},
\end{multline}
\begin{multline}
	\bm j_{(2+2)}^{v} = \frac{16\pi^2 e^2 v_0}{\omega^2}\sum_{\bm p}\expval{\Re[V_{cv}^*(\bm p)U_{c,\bm p';c,\bm p}^*U_{c,\bm p';v,\bm p}]\frac{[\bm E \cdot(\bm p+\bm p')](\bm p-\bm p')}{p^2}} \mathcal{D}_0(p)\dv{\epsilon(p)}{p}\delta(\varepsilon_{c,\bm p}-\varepsilon_{v,\bm p}-\omega)\tau^{(1)}_{\bm p}.
\end{multline}
As a result, the total correction is given by 
\begin{equation}\label{eq:ball_2_p_2_full}
	\bm j_{(2+2)}^{\text{ball}} = -\frac{16\pi^2 e^2 v_0}{\omega^2}\sum_{\bm p}\expval{\Re[V_{cv}^*(\bm p)U_{c,\bm p';c,\bm p}^*U_{c,\bm p';v,\bm p}]\frac{[\bm E \cdot(\bm p+\bm p')]\bm p'}{p^2}} \mathcal{D}_0(p)\dv{\epsilon(p)}{p}\delta(\varepsilon_{c,\bm p}-\varepsilon_{v,\bm p}-\omega)\tau^{(1)}_{\bm p}.
\end{equation}
For $\bm E \parallel x$ we obtain
\begin{equation}
	\text{Short-range potential:}\qquad	 j_{(2+2),x}^{\tau} = j_{(2+2),x}^{v}  = 0,\quad  j_{(2+2),x}^{\text{ball}}  = j_{(2+2),x}^{{\cal D}} = \frac{3}{16}\frac{e^3 E_x^2 \lambda }{v_0^3}\dv{\epsilon(p_\omega)}{p},
\end{equation} 
\begin{multline}
	\text{Coulomb potential:}\qquad	j_{(2+2),x}^{v} =-2 j_{(2+2),x}^{\tau} = \frac{\lambda e^3 E_x^2}{16 v_0^3}\dv{\epsilon(p_\omega)}{p},\\  j_{(2+2),x}^{{\cal D}} = \frac{9e^3 E_x^2 \lambda }{32v_0^3}\dv{\epsilon(p_\omega)}{p},\qquad  j_{(2+2),x}^{\text{ball}} = \frac{5}{16}\frac{e^3 E_x^2 \lambda }{v_0^3}\dv{\epsilon(p_\omega)}{p}.
\end{multline} 
Note that hereafter we assume $f_0\qty(\varepsilon_{c,\bm p})=0$, $f_0\qty(\varepsilon_{v,\bm p})=1$.

\subsection{Interference of the first-order and third-order processes}

\begin{figure}[htp]
	\centering
	\includegraphics[scale=0.4]{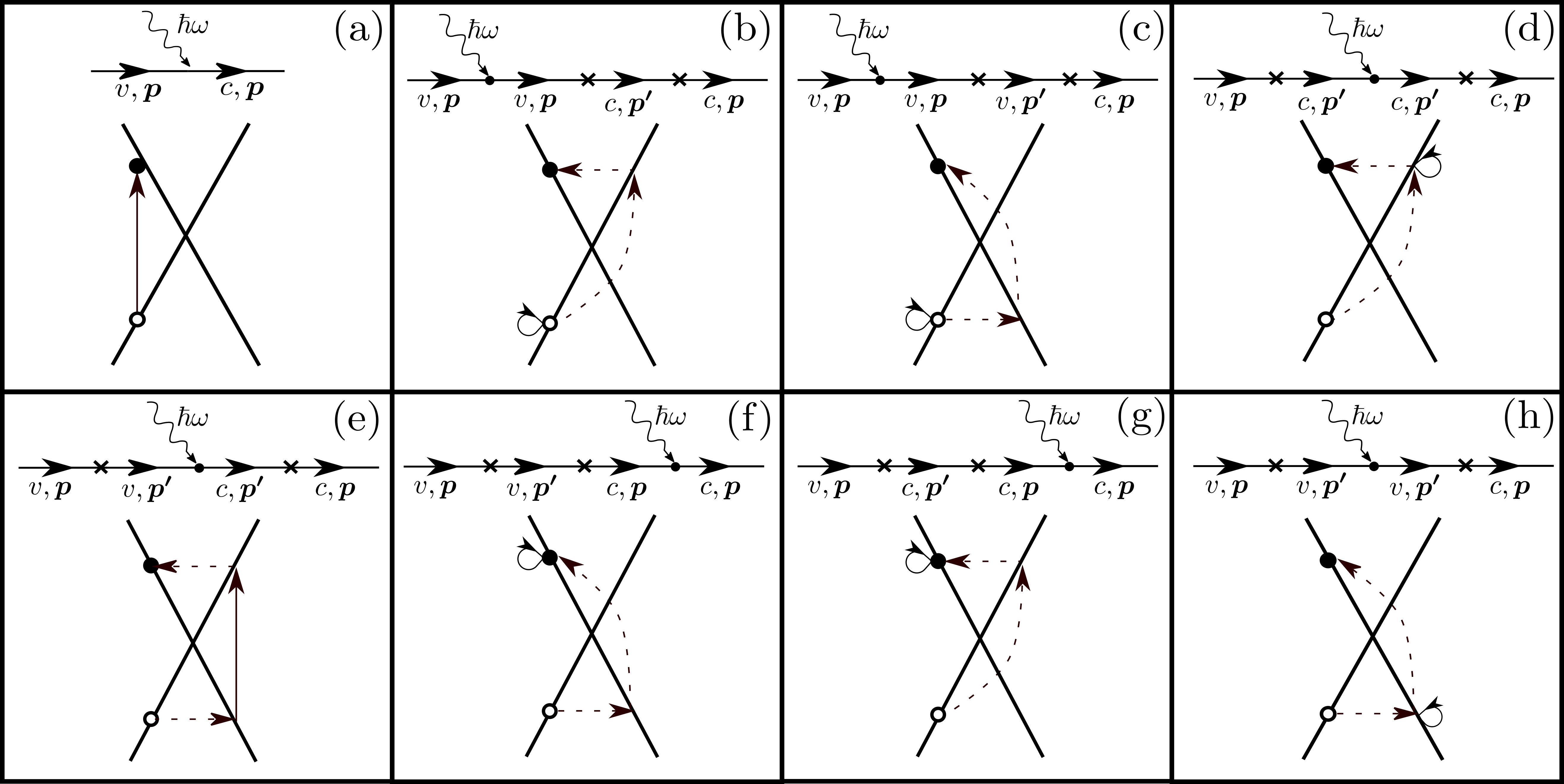} 
	\caption{First-order (a) and third-order (b)-(h) contributions to the matrix element of the transition $v,\bm p \rightarrow c,\bm p$.  
	%The solid lines denote optical transitions with photon absorption and dashed lines represent scattering. 
	Solid and dashed lines denote interaction of electrons with photons and impurities, respectively.}
	\label{fig:S1p3}
\end{figure}

A contribution to the photocurrent  in the first order of $\lambda$ also comes from the interference of the first-order process shown in
%Also the contribution to the photocurrent in the first order of $\lambda$ comes from the interference of the first-order
Fig.~\ref{fig:S1p3}(a) with the matrix element 
\begin{equation}\label{eq:3p1a}
	M^{(a)} = V_{cv}(\bm p)
\end{equation}
with one of the third-order processes shown in Fig.~\ref{fig:S1p3}(b)-(h). 

%\begin{equation}
%	\bm j_{(1+3)}^{\text{ball}} = e\sum_{\bm p}W_{c,\bm p;v,\bm p}(\bm v_{c,\bm p}\tau^{(1)}_{c,\bm p}-\bm v_{v,\bm p}\tau^{(1)}_{v,\bm p})
%\end{equation}

Similarly to the contribution of the second-order processes, the third-order processes contribute only if the energy conservation law holds for one of the intermediate states. This is not the case for processes shown in Fig.~\ref{fig:S1p3}(c) and Fig.~\ref{fig:S1p3}(g):
\begin{subequations}
\begin{equation}\label{eq:3p1c}
	M^{(c)}_{c,\bm p;v,\bm p}  = \frac{V_{vv}(\bm p)U_{v,\bm p';v,\bm p}U_{c,\bm p;v,\bm p'}}{\omega}\qty(\PV\frac{1}{\varepsilon_{c,\bm p}-\varepsilon_{v,\bm p'}}-i \pi \delta(\varepsilon_{c,\bm p}-\varepsilon_{v,\bm p'})).
\end{equation}
\begin{equation}\label{eq:3p1g}
	M^{(g)}_{c,\bm p;v,\bm p}  = -\frac{U_{c,\bm p';v,\bm p}U_{c,\bm p;c,\bm p'}V_{cc}(\bm p)}{\omega}\qty(\PV\frac{1}{\varepsilon_{v,\bm p}-\varepsilon_{c,\bm p'}}-i \pi \delta(\varepsilon_{v,\bm p}-\varepsilon_{c,\bm p'})).
\end{equation}
\end{subequations}
Since the first-order process~Fig.~\ref{fig:S1p3}(a) does not include $\delta$-function contribution, for the interference of processes (a) and (c) or (a) and (g) we need to take $\delta$-function from~\eqref{eq:3p1c} or~\eqref{eq:3p1g}, respectively, but %from the expression for the energy spectrum one can see that 
they are never satisfied because the energies in the valence and conduction bands are different. So, those processes do not contribute to the current for either symmetric or asymmetric spectrum.

The contribution from the process shown in Fig.~\ref{fig:S1p3}(e) is also absent. Indeed,
\begin{multline}
	M^{(e)}_{c,\bm p;v, \bm p} = \frac{U_{v,\bm p';c,\bm p}V_{cv}(\bm p')U_{c,\bm p;c,\bm p'}}{(\varepsilon_{v,\bm p}-\varepsilon_{v,\bm p'}+i0)(\varepsilon_{c,\bm p}-\varepsilon_{c,\bm p'}+i0)} =\\ =  -i \pi U_{v,\bm p';c,\bm p}V_{cv}(\bm p')U_{c,\bm p;c,\bm p'}\qty[\PV \frac{\delta(\varepsilon_{c,\bm p}-\varepsilon_{c,\bm p'})}{\varepsilon_{v,\bm p}-\varepsilon_{v,\bm p'}}+\PV\frac{\delta(\varepsilon_{v,\bm p}-\varepsilon_{v,\bm p'})}{\varepsilon_{c,\bm p}-\varepsilon_{c,\bm p'}}] =0,
\end{multline} 
since the spectrum is isotropic $\delta(\varepsilon_{(c,v),\bm p}-\varepsilon_{(c,v),\bm p'}) \propto \delta(p-p')$, and then
%\begin{equation*}
%	\delta(\varepsilon_{(c,v),\bm p}-\varepsilon_{(c,v),\bm p'}) \approx \delta(v_0 p-v_0 p')+\delta'(v_0 p-v_0 p')\dv{\epsilon(p)}{p}+{\cal O}\qty(\dv{\epsilon(p)}{p}),
%\end{equation*}
%\begin{equation*}
%		\frac{1}{\varepsilon_{(c,v),\bm p}-\varepsilon_{(c,v),\bm p'}+i0} = \pm \frac{1}{v_0 p-v_0 p'+i0},
%\end{equation*}
%\begin{equation*}
%	\delta(\varepsilon_{(c,v),\bm p}-\varepsilon_{(c,v),\bm p'}) \propto \delta(p-p'),\qquad \frac{1}{\varepsilon_{(c,v),\bm p}-\varepsilon_{(c,v),\bm p'}} = \pm \frac{1}{v_0 p-v_0 p'}
%\end{equation*}
%and 
\begin{equation*}
	 \PV  \int \dd{p}\int \dd{p'} \frac{\delta(\varepsilon_{(c,v),\bm p}-\varepsilon_{(c,v),\bm p'})}{\varepsilon_{(v,c),\bm p}-\varepsilon_{(v,c),\bm p'}} = \lim _{\delta\to 0}\Re\qty[\int \dd{p}\int \dd{p'} \frac{\delta(\varepsilon_{(c,v),\bm p}-\varepsilon_{(c,v),\bm p'})}{\varepsilon_{(v,c),\bm p}-\varepsilon_{(v,c),\bm p'}+i\delta}] =  \lim _{\delta\to 0} \Re\qty[\int \dd{p}\frac{1}{i\delta}] = 0.
\end{equation*}
%\begin{equation*}
%	 \PV  \int \dd{p}\int \dd{p'} \frac{\delta(p-p')}{v_0p-v_0p'} = \lim _{\delta\to 0}\Re\qty[\int \dd{p}\int \dd{p'}  \frac{\delta(p-p')}{v_0p-v_0p'+i\delta}] =  \lim _{\delta\to 0} \Re\qty[\int \dd{p}\frac{1}{i\delta}] = 0.
%\end{equation*}
Actually, we will show, that similarly to the interference of second order processes, the current from interference of the first and third order processes is zero in the case of the electron-hole symmetry. In this case the nonzero contributions from third order processes cancel each other: interference with (b) cancels by interference with (f), and (d) cancels by (h). Therefore we will consider them in pairs.

\begin{center}
\textbf{Third order processes (b) and (f)}
\end{center}

The matrix elements of processes are given by
\begin{subequations}
\begin{equation}
	M^{(b)}_{c,\bm p;v,\bm p} = \sum_{\bm p'}\frac{V_{vv}(\bm p) U_{c,\bm p';v,\bm p} U_{c,\bm p;c,\bm p'}}{ \omega(\varepsilon_{c,\bm p}-\varepsilon_{c,\bm p'}+i0)},
\end{equation}
\begin{equation}
	M^{(f)}_{c,\bm p;v,\bm p} =  -\sum_{\bm p'}\frac{U_{v,\bm p';v,\bm p}U_{c,\bm p; v,\bm p'}V_{cc}(\bm p)}{\omega(\varepsilon_{v,\bm p}-\varepsilon_{v,\bm p'}+i0)}.
\end{equation}
\end{subequations}
The generation rate for both processes reads
\begin{multline}
	G^{(b+f)}_{\bm p} = 4\pi\Re\qty[\qty(M^{(a)}_{c,\bm p;v,\bm p})^*(M^{(b)}_{c,\bm p;v,\bm p}+M^{(f)}_{c,\bm p;v,\bm p})]\delta(\varepsilon_{c,\bm p}-\varepsilon_{v,\bm p}-\omega)
	%(f_{v,\bm p}-f_{c,\bm p}) 
	=\\ = \frac{4\pi^2e}{\omega^2}\sum_{\bm p'}\Re[V_{cv}^*(\bm p) U_{c,\bm p' ; v,\bm p}U_{c,\bm p;c,\bm p'}][\bm E\cdot \bm v_{v,\bm p}\delta(\varepsilon_{c,\bm p}-\varepsilon_{c,\bm p'})+\bm E \cdot \bm v_{c,\bm p}\delta(\varepsilon_{v,\bm p}-\varepsilon_{v,\bm p'})]\delta(\varepsilon_{c,\bm p}-\varepsilon_{v,\bm p}-\omega),
	%(f_{v,\bm p}-f_{c,\bm p}) ,
\end{multline}
where we used the relations~\eqref{eq:U_rel}. One can see, that in the case of the electron-hole symmetry 
$$\bm E\cdot \bm v_{v,\bm p}\delta(\varepsilon_{c,\bm p}-\varepsilon_{c,\bm p'})+\bm E\cdot \bm v_{c,\bm p}\delta(\varepsilon_{v,\bm p}-\varepsilon_{v,\bm p'}) =0,$$
so we need to take into account asymmetry of the spectrum in the velocity or in the density of states:
\begin{subequations}
\begin{equation}
	\bm j_{(b+f)}^{{\cal D}} = \frac{16\pi^2 e^2 v_0^2}{\omega^2}\sum_{\bm p}\expval{\Re[V_{cv}^*(\bm p) U_{c,\bm p' ; v,\bm p}U_{c,\bm p;c,\bm p'}]}_{\bm p'}{\cal D}_0(p)\dv{\epsilon(p)}{p}\frac{(\bm E \cdot \bm p)\bm p}{p^2}\delta(\varepsilon_{c,\bm p}-\varepsilon_{v,\bm p}-\omega) \tau^{(1)}_{\bm p},
\end{equation}
\begin{equation}
	\bm j_{(b+f)}^{v} = \frac{16\pi^2 e^2v_0^2}{\omega^2}\sum_{\bm p}\expval{\Re[V_{cv}^*(\bm p) U_{c,\bm p' ; v,\bm p}U_{c,\bm p;c,\bm p'}]}_{\bm p'}{\cal D}_0(p)\dv{\epsilon(p)}{p}\frac{(\bm E \cdot \bm p)\bm p}{p^2}\delta(\varepsilon_{c,\bm p}-\varepsilon_{v,\bm p}-\omega) \tau^{(1)}_{\bm p}.
\end{equation}
\end{subequations}
Here $\expval{...}$ means angular averaging, and afterwards we put $p' = p$. One can see, that $\bm j_{(b+f)}^{{\cal D}} = \bm j_{(b+f)}^{v}$. Calculations yields 
\begin{equation}
	\text{Short-range:} \qquad \bm j_{(b+f)}^{{\cal D}}  =  \bm j_{(b+f)}^{v} = 0,
\end{equation}
\begin{equation}
	\text{Coulomb:} \qquad \bm j_{(b+f)}^{{\cal D}}  =  \bm j_{(b+f)}^{v} = -\frac{e^3\lambda E_x^2 }{16 v_0^3}\dv{\epsilon(p)}{p} \biggr|_{p=\omega/(2v_0)}.
\end{equation}

\begin{center}
\textbf{Third order processes (d) and (h)}
\end{center}
The matrix elements of these processes are given by 
\begin{subequations}
\begin{equation}
	M^{(d)}_{c,\bm p;v,\bm p} = \sum_{\bm p'}\frac{U_{c,\bm p';v,\bm p}V_{cc}(\bm p')U_{c,\bm p;c,\bm p'}}{(\varepsilon_{c,\bm p}-\varepsilon_{c,\bm p'}-\omega+i0)(\varepsilon_{c,\bm p}-\varepsilon_{c,\bm p'}+i0)},
\end{equation}
\begin{equation}
	M^{(h)}_{c,\bm p;v,\bm p} =\sum_{\bm p'} \frac{U_{v,\bm p';v,\bm p}V_{vv}(\bm p')U_{c,\bm p;v,\bm p'}}{(\varepsilon_{v,\bm p}-\varepsilon_{v,\bm p'}+i0)(\varepsilon_{c,\bm p}-\varepsilon_{v,\bm p'}+i0)}.
\end{equation}
\end{subequations}
Retaining the $\delta$-function from one of the denominators for each process we obtain the transition rate
\begin{equation}
	G_{\bm p}^{(d+h)} = -\frac{4\pi^2e}{\omega^2}\sum_{\bm p'}\Re[V_{cv}^*(\bm p) U_{c,\bm p' ; v,\bm p}U_{c,\bm p;c,\bm p'}][\bm E\cdot \bm v_{c,\bm p'}\delta(\varepsilon_{c,\bm p}-\varepsilon_{c,\bm p'})+\bm E \cdot \bm v_{v,\bm p'}\delta(\varepsilon_{v,\bm p}-\varepsilon_{v,\bm p'})]\delta(\varepsilon_{c,\bm p}-\varepsilon_{v,\bm p}-\omega).
\end{equation}
Similarly to the previous case, for the spectrum with electron-hole symmetry 
the term in square brackets is zero.
%$$[\bm E\cdot \bm v_{c,\bm p'}\delta(\varepsilon_{c,\bm p}-\varepsilon_{c,\bm p'})+\bm E\cdot \bm v_{v,\bm p'}\delta(\varepsilon_{v,\bm p}-\varepsilon_{v,\bm p'})] =0,$$
Therefore we need to take into account asymmetry of the spectrum in the velocity or in the density of states:
\begin{subequations}
\begin{equation}
	\bm j_{(d+h)}^{{\cal D}} = \frac{16\pi^2 e^2 v_0}{\omega^2}\sum_{\bm p}\expval{\Re[V_{cv}^*(\bm p) U_{c,\bm p' ; v,\bm p}U_{c,\bm p;c,\bm p'}]\frac{(\bm E \cdot \bm p')\bm p}{p^2}}_{\bm p'}{\cal D}_0(p)\dv{\varepsilon(p)}{p}\delta(\varepsilon_{c,\bm p}-\varepsilon_{v,\bm p}-\omega) \tau^{(1)}_{\bm p},
\end{equation}
\begin{equation}
	\bm j_{(d+h)}^{v} = -\frac{16\pi^2 e^2 v_0}{\omega^2}\sum_{\bm p}\expval{\Re[V_{cv}^*(\bm p) U_{c,\bm p' ; v,\bm p}U_{c,\bm p;c,\bm p'}]\frac{(\bm E \cdot \bm p')\bm p}{p^2}}_{\bm p'}{\cal D}_0(p)\dv{\varepsilon(p)}{p}\delta(\varepsilon_{c,\bm p}-\varepsilon_{v,\bm p}-\omega) \tau^{(1)}_{\bm p}.
\end{equation}
\end{subequations}
Those contributions are equal, but have opposite signs, hence they do not contribute to the current for both considered potentials.

\subsection{Total contribution}
We derived ballistic current for both short-range and Coulomb potentials. For the short-range potential the contribution comes only from the interference of the two second-order processes and the total contribution is equal to 
\begin{equation}
	\chi^{\text{ball, SR}} = \frac{3}{16}\frac{e^3 \lambda }{v_0^3}\dv{\epsilon(p)}{p}\biggr|_{p=\omega/(2v_0)}.
\end{equation}
In the case of the Coulomb potential we have contributions from interference of both two second-order processes and 
%the interference of 
first-order and third-order processes. They have different signs, but the latter is smaller. Their sum also reads
\begin{equation}
	\chi^{\text{ball, Coul}} = \frac{3}{16}\frac{e^3 \lambda }{v_0^3}\dv{\epsilon(p)}{p} \biggr|_{p=\omega/(2v_0)}.
\end{equation}

\section{Nonlinear ballistic current}
\label{sec:NonlinBall}

Here we consider the case of the short-range potential scattering only.
Kinetic equations for the occupation in the conduction band $\Delta f_{c\bm p}$ and a correction to the occupation in the valence band $1-\Delta f_{v\bm p}$ read
\begin{subequations}\label{eq:kineq}
\begin{equation}
\text{St}_c[\Delta f_{c\bm p}]= G_{\bm p}(1-\Delta f_{v\bm p}-\Delta f_{c\bm p}) +\sum_{\bm p'} \delta G_{\bm p \bm p'}(1-\Delta f_{v\bm p'}-\Delta f_{c\bm p}),
\end{equation}
\begin{equation}
\text{St}_v[\Delta f_{v\bm p}] = G_{\bm p}(1-\Delta f_{v\bm p}-\Delta f_{c\bm p}) +\sum_{\bm p'} \delta G_{\bm p' \bm p}(1-\Delta f_{v\bm p}-\Delta f_{c\bm p'}),
\end{equation}
\end{subequations}
where
\begin{equation}\label{eq:Gp}
G_{\bm p} = 2\pi \abs{V_{cv}(\bm p)}^2 \delta(2v_0p-\omega) = G_{-\bm p}
\end{equation}
is the rate of interband transitions in the absence of scattering and neglecting the warping, and $\delta G_{\bm p \bm p'}$ is a small generation rate $\bm p \leftarrow \bm p'$ yielding the ballistic photocurrent.

\subsection{Quadratic in intensity current}
%The occupations of the conduction and valence band are still described by the kinetic equations~\eqref{eq:kineq}, but we need to take into account difference in the harmonics relaxation times. 

The generation rate~\eqref{eq:Gp} consist of the zero and second harmonics so we define \begin{equation}
	G_{\bm p} = G_{\bm p}^{(0)}+G_{\bm p}^{(2)}.
\end{equation}
From the equations~\eqref{eq:kineq} we find the distribution functions linear in the light intensity:
%that the in the first order of the light intensity the distribution functions are given by
\begin{subequations}\label{Seq:dfcv}
\begin{equation}
	\Delta f_{c,\bm p} = \tau^c_1\sum_{\bm p'}\delta G_{\bm p,\bm p'}+\tau_\varepsilon^c G_{\bm p}^{(0)}+\tau_2^c G_{\bm p}^{(2)},
\end{equation} 
\begin{equation}
	\Delta f_{v,\bm p} = \tau_1^v\sum_{\bm p'}\delta G_{\bm p',\bm p}+\tau_\varepsilon^v G_{\bm p}^{(0)}+\tau_2^v G_{\bm p}^{(2)}.
\end{equation} 
\end{subequations}
Here we take into account difference in the relaxation times of the distribution function Fourier-harmonics.
Note that the first term here gives the photocurrent in the linear in the intensity regime, and the second and the third terms  play a role in the quadratic in the light intensity photocurrent.

Then we substitute $\Delta f_{c,v}$ back to the equation~\eqref{eq:kineq} and search for the first Fourier-harmonics 
%only because only it contribute 
contributing to the current:
\begin{subequations}\label{Seq:kin_eq_2}
\begin{equation}
	\frac{\delta f_{c,\bm p}}{\tau_1^c} = -G_{\bm p}\sum_{\bm p'}\qty(\tau_1^c\delta G_{\bm p,\bm p'}+\tau_1^v \delta G_{\bm p',\bm p})-\sum_{\bm p'}\delta G_{\bm p,\bm p'} \qty(\tau_\varepsilon^v G_{\bm p'}^{(0)}+\tau_2^v G_{\bm p'}^{(2)}+\tau_\varepsilon^c G_{\bm p}^{(0)}+\tau_2^v G_{\bm p}^{(2)}),
\end{equation}  
\begin{equation}
	\frac{\delta f_{v,\bm p}}{\tau_1^v} = -G_{\bm p}\sum_{\bm p'}\qty(\tau_1^c\delta G_{\bm p,\bm p'}+\tau_1^v \delta G_{\bm p',\bm p})-\sum_{\bm p'}\delta G_{\bm p',\bm p} \qty(\tau_\varepsilon^v G_{\bm p}^{(0)}+\tau_2^v G_{\bm p}^{(2)}+\tau_\varepsilon^c G_{\bm p'}^{(0)}+\tau_2^v G_{\bm p'}^{(2)}).
\end{equation}
\end{subequations}
The electron-hole asymmetry is taken in the first order:
\begin{equation}
	\Delta \tau_{1,2} = \frac{1}{v_0}\dv{\epsilon}{p}\overline{\tau}_{1,2},\qquad \delta G_{\bm p,\bm p'}^{\text{sym}} \equiv \frac{1}{v_0}\dv{\epsilon}{p} \delta G^S  
\end{equation}
Substituting it into the current we get
\begin{multline}
	\bm j_2^{\text{ball}} = -2v_0 e\qty(\frac{1}{v_0}\dv{\epsilon}{p})\sum_{\bm p,\bm p'} \frac{\bm p}{p}\left\{\overline{\tau}_1 \delta G^S\qty[\overline{\tau}_\varepsilon(G_{\bm p}^{(0)}+G_{\bm p'}^{(0)})+\overline{\tau}_2(G_{\bm p}^{(2)}+G_{\bm p'}^{(2)})+2G_{\bm p}\overline{\tau}_1]\right. + \\ +\left. \overline{\tau}_1 \delta G_{\bm p,\bm p'}^{\text{asym}} \qty[ 2 G_{\bm p} \overline{\tau}_1+\overline{\tau}_2( G_{\bm p}^{(2)}-G_{\bm p'}^{(2)})+2\overline{\tau}_\varepsilon(G_{\bm p}^{(0)}+G_{\bm p'}^{(0)})+ 2\overline{\tau}_2 (G_{\bm p}^{(2)}+G_{\bm p'}^{(2)})]\right\}.
\end{multline}
This yields the quadratic in the light intensity current in the form
\begin{equation}
	j_{2,x}^{\text{ball}} =  -\chi_1 E^2 \frac{\tau_1}{\tau_1^*}{F^2\over 24}\qty[\qty(18 \frac{\tau_1}{\tau_\varepsilon}-11\frac{\tau_2}{\tau_\varepsilon}+24)\cos(2\alpha)-\qty(12\frac{\tau_1}{\tau_\varepsilon}+\frac{\tau_2}{\tau_\varepsilon})\cos(4\alpha)],
\end{equation}

where $F=2eEv_0\tau_\varepsilon/\omega$.

\subsection{Saturation value of the current}
In the case of fast energy relaxation $\tau_\varepsilon \ll \tau_{1,2}^*$ we can solve the kinetic equations~\eqref{eq:kineq} for arbitrary light intensities  and obtain the photocurrent that contains all orders of light intensity.
%Let us consider the case of fast energy relaxation when
%\begin{equation}
%\text{St}_c[\Delta f_{c\bm p}] = {\Delta f_{c\bm p}\over \tau_\varepsilon^c},
%\qquad
%\text{St}_v[\Delta f_{v\bm p}] = {\Delta f_{v\bm p}\over \tau_\varepsilon^v}.
%\end{equation} 

We first find the corrections to the occupations at $\delta G_{\bm p \bm p'}=0$:
%Then we obtain the corrections to the occupations in the form
\begin{equation}
\Delta f_{v\bm p} = {\tau_\varepsilon^v\over \tau_\varepsilon^c}\Delta f_{c\bm p}, \qquad 
\Delta f_{c\bm p} ={G_{\bm p}\tau_\varepsilon^c \over 1+ 2\overline{\tau}_\varepsilon G_{\bm p}}, \qquad
\overline{\tau}_\varepsilon = {\tau_\varepsilon^c + \tau_\varepsilon^v\over 2}.
\end{equation}
Then we substitute these occupations to the rhs of the rate equations and find the current-carrying corrections to the distribution functions $\delta f_{c,v} \propto \delta G_{\bm p \bm p'}$:
\begin{equation}
\delta f_{c,v\bm p} = {\tau_\varepsilon^{c,v}\over 1 + 2\overline{\tau}_\varepsilon G_{\bm p}} \qty[A_{c,v} \pm G_{\bm p}\tau_\varepsilon^{v,c}(A_c-A_v)],
\end{equation}
where we introduced
\begin{subequations}
\begin{equation}
A_c = \sum_{\bm p'} \delta G_{\bm p \bm p'} \qty(1-{G_{\bm p'}\tau_\varepsilon^v \over 1+ 2\overline{\tau}_\varepsilon G_{\bm p'}}-{G_{\bm p}\tau_\varepsilon^c \over 1+ 2\overline{\tau}_\varepsilon G_{\bm p}}),
\end{equation}
\begin{equation}
A_v = \sum_{\bm p'} \delta G_{\bm p' \bm p} \qty(1-{G_{\bm p'}\tau_\varepsilon^c \over 1+ 2\overline{\tau}_\varepsilon G_{\bm p'}}-{G_{\bm p}\tau_\varepsilon^v \over 1+ 2\overline{\tau}_\varepsilon G_{\bm p}}).
\end{equation}
\end{subequations}		

In what follows we take equal relaxation times: $\tau_\varepsilon^c=\tau_\varepsilon^v=\tau$.
Then we obtain
\begin{equation}
\delta f_{c,v\bm p} = {\tau\over 1 + 2\tau G_{\bm p}} \qty[A_{c,v} \pm G_{\bm p}\tau(A_c-A_v)]
= {\tau\over 2} \qty[{A_c+A_v\over 1 + 2\tau G_{\bm p}} \pm (A_c-A_v)],
\end{equation}
where
\begin{equation}
{A_c \pm A_v \over 2} = \sum_{\bm p'} \delta G_{\bm p \bm p'}^{s,a}\qty(1-{G_{\bm p'}\tau \over 1+ 2\tau G_{\bm p'}}-{G_{\bm p}\tau \over 1+ 2\tau G_{\bm p}}),
\qquad
\delta G_{\bm p \bm p'}^{s,a} = {\delta G_{\bm p \bm p'} \pm \delta G_{\bm p' \bm p}\over 2}.
\end{equation}
This yields
\begin{equation}
\delta f_{c,v\bm p} = \tau \sum_{\bm p'} \qty(1-{G_{\bm p'}\tau \over 1+ 2\tau G_{\bm p'}}-{G_{\bm p}\tau \over 1+ 2\tau G_{\bm p}}) \qty({\delta G_{\bm p \bm p'}^{s}\over 1 + 2\tau G_{\bm p}} \pm\delta G_{\bm p \bm p'}^{a}).
\end{equation}
Note, that expansion of this function up to the second order of the light intensity will lead to the Eq.~\eqref{Seq:kin_eq_2} if we replace all relaxation times with $\tau_\varepsilon$ there.

The ballistic photocurrent density is calculated as follows
\begin{equation}
\bm j^{\rm ball} 
%= e\sum_{\bm p} \qty[\bm v_{c\bm p}\delta f_{c\bm p} + \bm v_{v\bm p}(1-\delta f_{v\bm p})]
= e\sum_{\bm p} \qty(\bm v_{c\bm p}\delta f_{c\bm p} - \bm v_{v\bm p} \delta f_{v\bm p})
.
\end{equation}
Substitution of $\delta f_{c,v}$ yields
\begin{equation}
\bm j^{\rm ball} = e\tau \sum_{\bm p, \bm p'}\qty(1-{G_{\bm p'}\tau \over 1+ 2\tau G_{\bm p'}}-{G_{\bm p}\tau \over 1+ 2\tau G_{\bm p}}) \qty[ \qty(\bm v_{c\bm p} - \bm v_{v\bm p}) {\delta G_{\bm p \bm p'}^{s}\over 1 + 2\tau G_{\bm p}} + \qty(\bm v_{c\bm p} + \bm v_{v\bm p}) \delta G_{\bm p \bm p'}^{a}].
\end{equation}
%
%\begin{multline}
%\bm j^{\rm ball} =e\sum_{\bm p} {\bm v_{c\bm p}\tau_\varepsilon^c \qty[A_{c} + G_{\bm p}\tau_\varepsilon^{v}(A_c-A_v)] - \bm v_{v\bm p}\tau_\varepsilon^v\qty[A_{v}- G_{\bm p}\tau_\varepsilon^{c}(A_c-A_v)]\over 1 + 2\overline{\tau}_\varepsilon G_{\bm p}}
%\\ 
%=e\sum_{\bm p} {\bm v_{c\bm p}\tau_\varepsilon^c A_{c}- \bm v_{v\bm p}\tau_\varepsilon^v A_{v} + \tau_\varepsilon^c\tau_\varepsilon^{v}(A_c-A_v)G_{\bm p}(\bm v_{c\bm p}+\bm v_{v\bm p}) \over 1 + 2\overline{\tau}_\varepsilon G_{\bm p}}
%.
%\end{multline}
%
Since 
%$\delta G_{\bm p \bm p'}^s \propto \dd \epsilon/\dd p$, and  
$\bm v_{c,v\bm p}= (\pm v_0  +  \dd \epsilon/\dd p)\bm n_{\bm p}$, where $\bm n_{\bm p}=\bm p/p$,
%in the first order in $ \dd \epsilon/\dd p$
we obtain
\begin{equation}
\bm j^{\rm ball} = e\tau \sum_{\bm p, \bm p'}\qty(1-{G_{\bm p'}\tau \over 1+ 2\tau G_{\bm p'}}-{G_{\bm p}\tau \over 1+ 2\tau G_{\bm p}}) 2\bm n_{\bm p} \qty( v_0 {\delta G_{\bm p \bm p'}^{s}\over 1 + 2\tau G_{\bm p}} + \dv{\epsilon}{p} \delta G_{\bm p \bm p'}^{a}).
\end{equation}

Let us analyze the limit of high intensity $G_{\bm p}\tau \to \infty$.
We see that $\delta G_{\bm p \bm p'}^{s}$ gives a smaller contribution than $\delta G_{\bm p \bm p'}^{a}$. Therefore we get
\begin{equation}
\bm j^{\rm ball}(I \to \infty) = 2e\tau\dv{\epsilon}{p_\omega} \sum_{\bm p, \bm p'}\qty(1-{G_{\bm p'}\tau \over 1+ 2\tau G_{\bm p'}}-{G_{\bm p}\tau \over 1+ 2\tau G_{\bm p}}) \bm n_{\bm p}  \delta G_{\bm p \bm p'}^{a}.
\end{equation}
Since the term in brackets is symmetric and $\delta G_{\bm p \bm p'}^{a}$ is asymmetric at an interchange $\bm p \leftrightarrow \bm p'$, we can rewrite
\begin{equation}
\bm j^{\rm ball}(I \to \infty) = e\tau\dv{\epsilon}{p_\omega} \sum_{\bm p, \bm p'}\qty(1-{G_{\bm p'}\tau \over 1+ 2\tau G_{\bm p'}}-{G_{\bm p}\tau \over 1+ 2\tau G_{\bm p}}) (\bm n_{\bm p}-\bm n_{\bm p'})  \delta G_{\bm p \bm p'}^{a}.
\end{equation}
Then changing $\bm p \leftrightarrow \bm p'$ in the 2nd term in the first brackets we obtain
\begin{equation}
\bm j^{\rm ball}(I \to \infty) = e\tau\dv{\epsilon}{p_\omega} \sum_{\bm p, \bm p'} {(\bm n_{\bm p}-\bm n_{\bm p'})  \delta G_{\bm p \bm p'}^{a} \over  1+ 2\tau G_{\bm p}}.
\end{equation}
%One can not neglect unity in the denominator because $G_{\bm p}$ equals to zero at a certain angle (where $\bm p \parallel \bm E$). In vicinity of this angle $\Delta\theta \sim 1/\sqrt{I}$ the integrand has a maximum giving the main contribution. This yields the asymptotics $j^{\rm ball}(I \to \infty) \propto \sqrt{I}$.

%We perform integration over absolute values of momenta as follows:
%\begin{equation}
%\sum_{\bm p, \bm p'} \mathcal F(\varepsilon_p, \bm n_{\bm p}, \bm n_{\bm p'}) = g^2(\omega/2) \int\limits_{-\infty}^\infty \dd \Delta \left< \mathcal F(\Delta, \bm n_{\bm p}, \bm n_{\bm p'}) \right>_{\bm n_{\bm p},\bm n_{\bm p'}},
%\end{equation}
%where $g(\varepsilon) \propto \varepsilon$ is the density of states and it is assumed that $\mathcal F$ has a sharp maximum at $\Delta\to 0$ where $\Delta=(\varepsilon_p-\omega/2)\tau$.
%Since 
%\begin{equation}
%\delta G_{\bm p \bm p'}^{a} = {C(\bm n_{\bm p}, \bm n_{\bm p'})\delta(\varepsilon_p-\varepsilon_{p'})\over \Delta^2 + 1}, \qquad
%G_{\bm p} = {A(\bm n_{\bm p})\over \Delta^2 +1}, \qquad {1\over  1+ 2\tau G_{\bm p}} = {\Delta^2 + 1\over \Delta^2 + 1 + 2\tau A(\bm n_{\bm p})},
%\end{equation}
%where $\Delta=2(\varepsilon_p-\omega/2)\tau$,
%we have
Calculation gives 
\begin{equation}
\bm j^{\rm ball}(I \to \infty) 
%e\tau\dv{\epsilon}{p_\omega} g(\omega)  \int\limits_{-\infty}^\infty \dd \Delta \left< {(\bm n_{\bm p}-\bm n_{\bm p'}) C(\bm n_{\bm p}, \bm n_{\bm p'}) \delta(p-p')\over \Delta^2 + 1 + 2\tau A(\bm n_{\bm p})}\right>_{\bm p,\bm p'}
 = \frac{4}{3} \tilde{\chi}_1^{\rm ball} E^2 \expval{\frac{(\bm n_{\bm p}-\bm n_{\bm p'})G^A(\theta_{\bm p},\theta_{\bm p'},\alpha)}{\sqrt{1+F^2 \sin^2(\theta_{\bm p}-\alpha)}}}_{\bm p,\bm p'},
\end{equation}
where the angular averaging is performed at $p = p'$, and 
\begin{multline}
	G^{A}(\theta_{\bm p},\theta_{\bm p'},\alpha) = - 2 \cos^2\qty(\theta_{\bm p}-\theta_{\bm p'}\over 2) \cos\qty(\alpha-{\theta_{\bm p}+\theta_{\bm p'}\over 2})\sin\qty(\theta_{\bm p}-\theta_{\bm p'}\over 2) \times \\ \times [\sin(\alpha+2\theta_{\bm p})+\sin(\alpha+2\theta_{\bm p'})-2\sin(\alpha-2(\theta_{\bm p}+\theta_{\bm p'}))].
\end{multline}
For the $x$ component of the current we have 
\begin{equation}
	j_x^{\rm ball}(I \to \infty) = \frac{4}{3} \tilde{\chi}_1^{\rm ball} E^2 \expval{\frac{\qty[\cos(\theta_{\bm  p})-\cos(\theta_{\bm  p'})] G^A(\theta_{\bm p},\theta_{\bm p'},\alpha)}{\sqrt{1+F^2 \sin^2(\theta_{\bm p}-\alpha)}}}.
\end{equation}
Substituting $\theta_{\bm p,\bm p'}\to\theta_{\bm p,\bm p'}+\alpha$ we obtain 
\begin{equation}
	j_x^{\rm ball}(I \to \infty) = \frac{2}{3} \tilde{\chi}_1^{\rm ball} E^2 \expval{\frac{\cos\qty(\theta_{\bm p}+\theta_{\bm p'}\over 2)\sin^2(\theta_{\bm p}-\theta_{\bm p'})}{\sqrt{1+F^2 \sin^2(\theta_{\bm p})}} \qty[{\cal A}(\theta_{\bm p},\theta_{\bm p'})\cos(2\alpha)-{\cal B}(\theta_{\bm p},\theta_{\bm p'})\cos(4\alpha)]},
\end{equation}
where 
\begin{equation}
	{\cal A}(\theta_{\bm p},\theta_{\bm p'}) = \cos\qty(3 \theta_{\bm p}-\theta_{\bm p'}\over 2)+\cos\qty(3 \theta_{\bm p'}-\theta_{\bm p}\over 2)+2\cos\qty(3(\theta_{\bm p}+\theta_{\bm p'})\over 2),
\end{equation}
\begin{equation}
	{\cal B}(\theta_{\bm p},\theta_{\bm p'}) = \cos\qty(5\theta_{\bm p}+\theta_{\bm p'}\over 2)+\cos\qty(5\theta_{\bm p'}+\theta_{\bm p}\over 2)+2\cos\qty(5(\theta_{\bm p}+\theta_{\bm p'})\over 2).
\end{equation}
Performing angular averaging over $\theta_{\bm p}$, $\theta_{\bm p'}$ and leaving the leading in the limit $F\to \infty$ terms only, we obtain
%\begin{equation}
%	j_x^{\rm ball}(I \to \infty) = \tilde{\chi}_1^{\rm ball} E^2  \frac{2(1+F^2)(16+5F^2)\text{E}(m)-(32+26F^2+3F^4)\text{K}(m)}{18 F^4\sqrt{1+F^2}\pi}(\cos(2\alpha)-\cos(4\alpha)).
%\end{equation}
%In the limit $F\to \infty$
\begin{equation}
	j_x^{\rm ball}(I \to \infty) 
%	= \tilde{\chi}_1^{\rm ball} E^2  \times \frac{10-\ln(64)-3\ln(F)}{18 F \pi}(\cos(2\alpha)-\cos(4\alpha))
 \approx  -\tilde{\chi}_1^{\rm ball} E^2  \frac{\ln F}{6 \pi F}(\cos2\alpha-\cos4\alpha).
\end{equation}

\section{Linear and nonlinear shift current}
The shift photocurrent is given by 
\begin{equation}\label{eq:shift_cur}
	\bm j = e \sum_{\bm p} G_{\bm p}(f_{v,\bm p}-f_{c,\bm p})\bm R_{cv}(\bm p),
\end{equation}
where the shift vector  
\begin{equation}\label{Seq:R_def}
\bm R_{cv}(\bm p)=-\bm\nabla_{\bm p}\text{arg}\qty(V_{cv})+\bm \Omega_c(\bm p) - \bm \Omega_v(\bm p)
\end{equation} 
with $V_{cv}(\bm p)$ being the matrix element~\eqref{Seq:mat_el_1}, and $\bm \Omega_{c,v}$ are the Berry curvatures of the conduction and valence bands $\bm \Omega_{n}=i\hbar\bra{\psi_n}\bm \nabla_{\bm p}\ket{\psi_n}$ with $\psi_{c,v}$  given by Eq.~\eqref{eq:S:psi_c_v}. Calculation yields 
%for the system under study is given by 
\begin{equation}\label{Seq:R_val}
	\bm R_{cv}(\bm p) = \frac{6 p \lambda}{v_0}\frac{\sin(\alpha+\theta_{\bm p})\bm e_x+\cos(\alpha+\theta_{\bm p})\bm e_y}{\sin(\alpha-\theta_{\bm p})}.
\end{equation}

\subsection{Linear shift current}
In the linear in the light intensity regime we 
%assume unoccupied bands 
take in Eq.~\eqref{eq:shift_cur}
$f_{v,\bm p}-f_{c,\bm p} = 1$. Then the calculation of the photocurrent gives 
\begin{equation}\label{Seq:jsh_lin}
	 j_x^{\text{sh}}+ij_y^{\text{sh}}  = \chi_1^{\text{sh}}E^2e^{-2i\alpha}, \qquad 
	 \chi_1^{\text{sh}}=-\frac{3}{8}\frac{e^3 E^2 \lambda}{v_0^2}.
\end{equation}
The difference in $\chi_1^{\text{sh}}$ with the result of Ref.~\cite{Kim_Shift_2017} in the factor $-2$ is due to two reasons: (i) the factor 2 is coming from the fact, that the factor 2 need to be included into the phenomenological equation for the current ($J^{\text{shift}}_j = 2\sum_{i=x,y} \chi_j^{ii} E_i E_i$ before Eq.(2) in~\cite{Kim_Shift_2017}) similarly to Ref.~\cite{Sipe_SOR_2000} or directly into Eq.(2) in~\cite{Kim_Shift_2017} similarly to this paper; (ii) the minus sign comes from the different definition of the shift vector  in Ref.~\cite{Kim_Shift_2017} (as well as in Ref.~\cite{Sipe_SOR_2000}) with the conventional one, Eq.~\eqref{Seq:R_def}, see, e.g. Ref.~\cite{Belinicher_Ivchenko_Sturman_1982}.

\subsection{Quadratic in intensity current}
Since for the shift current we take into account only direct optical transitions, for calculations of the quadratic in the intensity current we need to substitute in~\eqref{eq:shift_cur} the second order distribution function~\eqref{Seq:dfcv} setting $\delta G_{\bm p,\bm p'} = \delta G_{\bm p',\bm p} = 0$ without taking into account warping in $G_{\bm p}$ because it is taken into account in the shift vector~\eqref{Seq:R_val}.

From the calculation we obtain the correction to the answer~\eqref{Seq:jsh_lin} in the order of $E^4$
\begin{equation}\label{eq:sec_or_shift}
	j_{2,x}^{\text{sh}}=\frac{3}{8}\frac{e^5 E^4\lambda \tau_\varepsilon(2\tau_\varepsilon+\tau_2)}{\omega^2}\cos(2\alpha)\equiv 
	%\chi_1^{\text{sh}}E^2\frac{\tau_\varepsilon}{2}\qty(e E v_0\over \omega)^2(2\tau_\varepsilon+\tau_2)\cos(2\alpha) = 
	 \chi_1^{\text{sh}}E^2\frac{F^2}{8}\qty(2+\frac{\tau_2}{\tau_\varepsilon})\cos(2\alpha).
\end{equation}
Comparing to the case of the ballistic current, one can see  that the terms $E_{\pm}^4$ are absent in the shift current.

\subsection{Shift current for the arbitrary intensity}
The theory of the nonlinear optical absorption in topological insulators in all orders of the light intensity was developed in the Ref.~\cite{Nonlin_our_PRB_2022}. The solution of the kinetic equation that takes into account elastic and inelastic scattering with the scattering times $\tau_p$ and $\tau_\varepsilon$, respectively, for the linearly polarized excitation light at the arbitrary intensity is given by 
\begin{equation}\label{Seq:Nf}
	f_{v,\bm p}-f_{c,\bm p} = \frac{1}{[1+2G(\bm p)\tau]\qty[1+\Psi_{\text{lin}}(p)(\tau_\varepsilon/\tau-1)]},
\end{equation}
where $\tau^{-1} = \tau_p^{-1}+\tau_\varepsilon^{-1}$  and we blurred the $\delta$-function in $G(\bm p)$:  
\begin{equation}
	G(\bm p) = \frac{2\abs{V_{cv}(\bm p)}^2/\tau}{(\varepsilon_{c,\bm p}-\varepsilon_{v,\bm p}-\omega)^2+(1/\tau)^2},
\end{equation}
and 
\begin{equation}
	\Psi_{\text{lin}}(p) = \expval{G(\bm p)\over G(\bm p)+1/(2\tau)}_{\theta_{\bm p}} = 1-\sqrt{\frac{1+\Delta^2}{1+2{\cal E}^2+\Delta^2}},
\end{equation}
where ${\cal E} = \sqrt{2}ev_0 E \tau /\omega$ is dimensionless electric field amplitude defined in~\cite{Nonlin_our_PRB_2022} ($\mathcal E= F/\sqrt{2}$) and $\Delta=(2v_0p-\omega)\tau$.
Substituting~\eqref{Seq:Nf} into~\eqref{eq:shift_cur} we obtain 
\begin{equation}\label{eq:Nonlin}
	\bm j^{sh} = -\frac{e\omega}{2\pi \hbar^2 v_0^2}\int\limits_0^\infty \dd{\Delta}\frac{\bm \Phi(\Delta)}{1+\Psi(\Delta)\tau_\varepsilon/\tau_p},
\end{equation}
where
\begin{equation}
	\bm \Phi(\Delta) = \expval{\frac{\bm R_{cv}(\bm p)}{\Delta^2+1+4\abs{V_{cv}(\bm p)}^2\tau^2}}_{\theta_{\bm p}}.
\end{equation}  
Calculation yields 
\begin{equation}
		\bm j^{\text{sh}} = -\frac{3e^3 \lambda {\cal F} E_0^2}{8\hbar^3v_0^2} \frac{2}{\pi{\cal E}^2}\int\limits_0^\infty \dd{\Delta} \frac{\Psi_{\text{lin}}(\Delta)}{1+\Psi_{\text{lin}}(\Delta)\tau_\varepsilon/\tau_p} \times \qty[\cos(2\alpha)\bm e_x-\sin(2\alpha)\bm e_y].
	\end{equation}
	According to Ref.~\cite{Nonlin_our_PRB_2022} the nonlinear light absorption $\eta_{\text{lin}}({\cal E},\tau_\varepsilon/\tau_p)$ is given by
	\begin{equation}
	\label{eq:S:eta}
		\frac{\eta_{\text{lin}}({\cal E},\tau_\varepsilon/\tau_p)}{\eta_0} =\frac{2}{\pi{\cal E}^2} \int\limits_0^\infty \dd{\Delta} \frac{\Psi_{\text{lin}}(\Delta)}{1+\Psi_{\text{lin}}(\Delta)\tau_\varepsilon/\tau_p},
	\end{equation}
	where $\eta_0 = \pi e^2 /(4 c)$. Therefore we see that the nonlinear photocurrent could be represent through the nonlinear light absorption: 
	%studied in the paper~\cite{Nonlin_our_PRB_2022}:
	\begin{equation}\label{Seq:j_nonlin}
		\bm j^{\text{sh}}({\cal E},\tau_\varepsilon/\tau_p)  = \chi_1 E_0^2 \times \frac{\eta_{\text{lin}}({\cal E},\tau_\varepsilon/\tau_p)}{\eta_0} \times \qty[\cos(2\alpha)\bm e_x-\sin(2\alpha)\bm e_y].
	\end{equation}

\subsection{Effective inelastic scattering}

In the case of the effective inelastic scattering $\tau_\varepsilon/\tau_p \to 0$ we obtain from Eqs.~\eqref{eq:S:eta},~\eqref{Seq:j_nonlin} 
\begin{equation}
\chi_1^{\rm shift}(E) =  -{3 \lambda e^3 \over 2\pi v_0^2} \times {\text{K}(m)-(1+F^2)\text E(m)\over F^2 \sqrt{1+F^2}},
\end{equation}
where $m=F^2/(1+F^2)$, and 
$\text E(m)[\text K(m)] = \int_0^{\pi/2} \dd \theta (1 - m \sin^2\theta)^{\pm 1/2}$ are the complete elliptic integrals.
Expanding this result at small values of $F$ we get
\begin{equation}
	{\chi_1^{\rm shift}(E)\over \chi_1^{\rm shift}} \approx 1-\frac{3F^2}{8}+{\cal O}\qty(F^4).
\end{equation}
This expression also follows from the limit $\tau_2 \approx \tau_\varepsilon$ in the Eq.~\eqref{eq:sec_or_shift}:
\begin{equation}
	1-\frac{F^2}{8}\qty(2+\frac{\tau_2}{\tau_\varepsilon}) \approx [\tau_2 \approx \tau_\varepsilon] \approx 1-\frac{3F^2}{8}.
\end{equation}

	In the opposite limit for $F\to \infty$ we obtain 
	\begin{equation}
		{\chi_1^{\rm shift}(E)\over \chi_1^{\rm shift}}  \xrightarrow{F\to \infty} \frac{4}{\pi F}.
	\end{equation}
	
%----------------------------------
\renewcommand{\i}{\ifr}
\let\oldaddcontentsline\addcontentsline% Store \addcontentsline
\renewcommand{\addcontentsline}[3]{}% Make \addcontentsline a no-op

\bibliography{SI_8_NL.bbl}

%\bibliography{citations}% Produces the bibliography via BibTeX.